\pdfoutput=1

\documentclass[11pt,review,authoryear]{elsarticle}

\usepackage{amsfonts}
\usepackage{amsmath}
\usepackage{hyperref}
\usepackage{enumitem}
\usepackage[utf8]{inputenc}
\usepackage{bm}
\usepackage{subfigure}
\usepackage{xcolor}
\usepackage{soul}
\usepackage[toc,page]{appendix}
\usepackage{tikz}
\usetikzlibrary{shapes,arrows}

\usepackage[margin=1.5cm]{geometry}

\biboptions{authoryear}

\newcommand\copyrighttext{%
  \footnotesize \copyright\ 2021. This manuscript version is made available under the CC-BY-NC-ND 4.0 license \url{http://creativecommons.org/licenses/by-nc-nd/4.0/} This work has been submitted to Collective Dynamics for possible publication. Copyright may be transferred without notice, after which this version may no longer be accessible.}
\newcommand\copyrightnotice{%
\begin{tikzpicture}[remember picture,overlay]
\node[anchor=south,yshift=160pt] at (current page.south) {\fbox{\parbox{\dimexpr\textwidth-\fboxsep-\fboxrule\relax}{\copyrighttext}}};
\end{tikzpicture}%
}

\makeatletter
\def\ps@pprintTitle{%
 \let\@oddhead\@empty
 \let\@evenhead\@empty
 \def\@oddfoot{}%
 \let\@evenfoot\@oddfoot}
\makeatother

\DeclareMathAlphabet{\mathcal}{OMS}{cmsy}{m}{n}

\begin{document}

\begin{frontmatter}

\title{Minimization of mean-CVaR evacuation time of a crowd using rescue guides: a scenario-based approach}

\author[1]{Anton von Schantz}\corref{cor1}
\ead{anton.von.schantz@aalto.fi}
\cortext[cor1]{Corresponding author}
\author[1]{Harri Ehtamo}
\ead{harri.ehtamo@aalto.fi}
\author[2]{Simo Hostikka}
\ead{simo.hostikka@aalto.fi}

\address[1]{Aalto University, Department of Mathematics and Systems Analysis, P.O. Box 11100, FI-00076 Aalto, Finland}
\address[2]{Aalto University, Department of Civil Engineering, P.O. Box 12100, FI-00076 Aalto, Finland}


%

\begin{abstract}
In case of a threat in a public space, the crowd in it should be moved to a shelter or evacuated without delays. Risk management and evacuation planning in public spaces should also consider uncertainties in the traffic patterns of crowd flow. One way to account for the uncertainties is to make use of safety staff, or guides, that lead the crowd out of the building according to an evacuation plan. Nevertheless, solving the minimum time evacuation plan is a computationally demanding problem. In this paper, we model the evacuating crowd and guides as a multi-agent system with the social force model. To represent uncertainty, we construct probabilistic scenarios. The evacuation plan should work well both on average and also for the worst-performing scenarios. Thus, we formulate the problem as a bi-objective scenario optimization problem, where the mean and conditional value-at-risk (CVaR) of the evacuation time are objectives. A solution procedure combining numerical simulation and genetic algorithm is presented. We apply it to the evacuation of a fictional passenger terminal. In the mean-optimal solution, guides are assigned to lead the crowd to the nearest exits. In contrast, in the CVaR-optimal solution, the focus is on solving the worst-case scenario's physical congestion. With one guide positioned behind each agent group near each exit, a plan that minimizes both objectives is obtained.
\end{abstract}

\end{frontmatter}
%

\copyrightnotice

\newpage

\section{Introduction}
\label{sec:intro}

In case of a safety threat in a public place, the crowd has to be evacuated fast. There are many studies where the optimal evacuation routes have been calculated \citep{haghani}. However, these theoretical solutions are not accompanied by practical solutions to implement and enforce these routes. An extreme suggestion is to send the routes to crowd members, to their cellphones \citep{optimizedroute}. Nevertheless, it is known that in emergencies, people tend to follow clear orders and guidance given by authorities, like security staff, and the use of security staff improves evacuation efficiency \citep{gwynne2016, proulx2002}. Thus, the coordination of security staff, or guides, is a more practical problem to solve.

The evacuation time of a crowd is sensitive to changes in conditions, even to deviations in individuals' positions \citep{safetyscience, latticegassimulation}. A working evacuation plan should be robust against them. Related to this, we recently raised the question in \citep{ourejor}: how should a crowd be evacuated when there is the possibility that a large part of it deviates from its usual rules of motion? Here, our objective is to study how the guides should be coordinated so that the crowd is evacuated in minimum time considering different crowd flow traffic scenarios. We are also interested in how the evacuation plan changes when the number of guides varies; e.g., some of them can be needed simultaneously in other operations. We will present our study in the context of a passenger terminal, even though other spaces containing a crowd could be used. Passenger terminals are characterized by high and fluctuating crowd flow, with a wide variety of people with different destinations to reach \citep{schultzandfricke, ali2019impact}. Also, they typically have security staff that can respond fast and guides the evacuation if needed.

When modeling the evacuation of a crowd using guides, it is appropriate to use the microscopic scale to focus directly on individuals and one-to-one interactions \citep{bellomo2012}. The social force model \citep{socialforce95} and cellular automaton model \citep{cellularautomaton} are the most well-known microscopic model types. Here, we use the social force model. In it, a mixture of socio-psychological and physical forces is assumed to influence an agent's motion in the crowd, which is modeled with Newtonian mechanics. When guide agents are added to the crowd, they also influence the motion of the other agents. In the original, stochastic version of the model, a Gaussian random force is added to an agent's equation of motion to describe intentional or unintentional deviations from the usual rules of motion.

Different aspects of the optimal use of guides in an evacuation have been studied. The main focus has been on the optimal proportion or number of guides \citep{trainedleader, evacuationassistants, optimizingproportion}. Choosing the number of guides alone does not guarantee a fast evacuation. If the guides' routes are not simultaneously optimized, it can even make the evacuation slower \citep{necessityguides}. The studies are mainly comparative, where different configurations are numerically simulated and compared. There are only a few studies where mathematical optimization has been used \citep{invisiblecontrol, maximumcoverage2, ourejor}. On the other hand, the need for rigorous mathematical approaches has been recognized \citep{haghani}.

In our recent study \citep{ourejor}, the number of guides, their initial positions, and exit assignments needed to minimize the crowd evacuation time is solved in a single optimization problem. In it, the crowd is modeled with the stochastic social force model to which interaction rules between regular agents and guides are added. The minimum time crowd evacuation problem is first formulated as a stochastic optimization problem. Then it is reformulated as a scenario optimization problem and solved with a combined numerical simulation and genetic algorithm (GA) \citep{ourejor}.

In a real-life evacuation, people can participate in various activities that do not immediately get them to the exit, which is a typical concern in evacuation modeling \citep{gwynne2016}. While these behavioral deviations could be implemented in the social force model framework, we assume them to be small compared to the deterministic part of a moving crowd controlled by guides and approximated with the Gaussian random force term. This approach does not suffice if a larger part of the crowd behaves differently. Instead, the different crowd flow traffic patterns could be modeled by changing some of the input parameters.

Robust optimization \citep{robustsurvey} or stochastic programming \citep{birge1982} can be used to deal with the uncertainty of the input parameters. In robust optimization, the input parameters are defined to belong to an uncertainty set, but no probabilities are assigned. Typically, a solution that optimizes the worst-case scenario is solved. However, this approach might result in an overly conservative evacuation plan. On the other hand, in stochastic programming, different input parameters' realizations are modeled as probabilistic scenarios. The scenario probabilities in passenger terminals could be obtained from passenger data or security cameras. Typically, the mean of the objective function is optimized, which is the mean of the evacuation time over the scenarios. This approach is appropriate when the optimal evacuation plan works well across all scenarios. The slowest scenarios can be accounted for by minimizing a risk measure like the conditional value-at-risk (${\textrm{CVaR}}_{\alpha}$) evacuation time. It gives the conditional mean evacuation time for the $1-\alpha$ percentage of the slowest scenarios \citep{valueatrisk}. Note that in the special case that the worst-case scenario probability equals or exceeds $1-\alpha$, the worst-case scenario's evacuation time equals $\mathrm{CVaR}_{\alpha}$ evacuation time. This will be the situation in the toy example of our paper.

Instead of deciding whether the mean or ${\textrm{CVaR}}_{\alpha}$ is a more appropriate objective function, the minimum time crowd evacuation problem can be formulated as a bi-objective problem similar to that in portfolio optimization \citep{valueatrisk}. Choosing optimal guides' routes is essentially a combinatorial optimization problem. Thus usual derivative-based methods cannot be applied. Instead, a combined numerical simulation and GA approach is used. The GA iteratively searches for the optimal solutions. At the same time, numerical simulation evaluates the objective function's values in different scenarios. A special kind of GA is needed for bi-objective problems. In this paper, we will use NSGA-II \citep{deb2002fast}.

In this paper, we model the evacuating crowd of passengers and guides with a deterministic version of the social force model from \citep{ourejor}. When a guide is within an interaction range from a passenger, it starts to follow the guide. Different crowd flow traffic patterns are modeled as probabilistic scenarios. Our main contribution is that we formulate a mean-${\mathrm{CVaR}}_{\alpha}$ crowd evacuation problem with rescue guides. The optimization variables are the initial positions and exit assignments of guides. Then, we present a solution procedure combining numerical simulation and NSGA-II. 

Our paper is structured as follows. In Sec.~\ref{sec:dynamics} the mathematical details of the crowd movement model are presented. In Sec.~\ref{sec:optimizationframework} we define the optimization model. In Sec.~\ref{sec:solutionmethod} we present the solution procedure and its implementation details. In Sec.~\ref{sec:problemsetting} we present the case study of an evacuation of a passenger terminal. The numerical results and their analysis are presented in Sec.~\ref{sec:numericalresults}. Finally, Sec.~\ref{sec:conclusion} is for discussion and conclusion.

\section{Evacuation model with guides}\label{sec:dynamics}

Next, we will present the crowd movement model used for the evacuation of a passenger terminal. The model is the deterministic version of the social force model presented in \citep{ourejor}. It is essentially Helbing's original social force model, with a velocity-dependent social repulsion force. The social force model has been extensively discussed in previous research. We refer the reader to \citep{ourejor} and its appendix for a detailed mathematical description of the individual force terms and parameter values used.

Let us start by denoting the index set of passenger agents, or passengers, with $N=\left\lbrace 1,...,n \right\rbrace$, the index set of guide agents, or guides, with $G=\left\lbrace n+1,...,n+m \right\rbrace$, their combined index set with $I=N \cup G$, the building space with $\Omega \subset \mathbb{R}^2$, the exits with $\mathcal{E} \subset \Omega$, and the walls with $W \subset \Omega$. The points associated with an exit we denote by $\varepsilon \subset \mathcal{E}$, so that $\mathcal{E} = \cup \varepsilon$. And, the points associated with a wall segment we denote by $w \subset W$, so that $W = \cup w$.

An agent $i \in I$ is circle-shaped with radius $r_i$ and mass $m_i$. At time $t$, its center of mass is $\mathbf{x}_i(t)$ and its velocity vector is $\mathbf{v}_i(t)$. Its change of position is then:

\begin{equation}\label{eq:changeofvelocity}
    \frac{d\mathbf{x}_i}{dt} = \mathbf{v}_i.
\end{equation}

Initially, all passengers $l \in N$ are heading towards their destination exit ${\varepsilon}_l^{des} \in \mathcal{E}$. The evacuation planner has instructed the guides $g \in G$ to move towards the exits ${\varepsilon}_g \in \mathcal{E}$. If a passenger comes within the interaction range $r_{guide}$ of a guide, the passenger starts to follow the guide. If the passenger is within the $r_{guide}$ range to two or more guides, it follows the closest one. Once it has started to follow a guide, it will not start to follow another. Let us define a binary variable:

\begin{equation}
    u_{lg}(t) = \left\{\begin{array}{ll}
        1, & \textrm{if passenger $l$ follows guide $g$ at time $t$}, \\
        0, & \textrm{otherwise},
    \end{array}
    \right.
\end{equation}

\noindent
for which it initially holds.

The change of velocity at time $t$ for passenger $l$ is given by the equation of motion:

\begin{equation}\label{eq:eqofmotion_a}
    m_l \frac{d\mathbf{v}_l}{dt} = (1-\sum_g u_{lg}) \mathbf{f}_{l}^0 + \sum_g u_{lg} \mathbf{f}_{lg} + \sum_i \mathbf{f}_{li} + \sum_{w \subset W} \mathbf{f}_{lw}.
\end{equation}

\noindent
Here, the term $\mathbf{f}_l^0$ is the driving force of agent $i$ to move towards its destination exit. It describes the attempt of agent $l$ to change its actual velocity $\mathbf{v}_l$ to a desired velocity $v_l^0 \mathbf{e}_l^{des}$ with a certain characteristic reaction time $\tau$:

\begin{equation}
    \mathbf{f}_{l}^0 = m_l \frac{v_l^0 \mathbf{e}_{l}^{des} (\mathbf{x}_l) - \mathbf{v}_l}{\tau}.
\end{equation}

\noindent
Here, $v_l^0$ is the desired speed of agent $l$. For simplicity, we assume it to be constant throughout the evacuation. In position $\mathbf{x} \in \Omega \setminus \left\lbrace {\varepsilon}_l^{des} \cup W \right\rbrace$, the unit vector $\mathbf{e}_l^{des}$ gives the direction of the shortest path of agent $l$ towards its destination exit ${\varepsilon}_l^{des}$:

\begin{equation}\label{eq:normedgradient}
    \mathbf{e}_l^{des} (\mathbf{x}) = -\frac{\nabla {D}_l^{des} (\mathbf{x})}{\lVert \nabla {D}_l^{des} (\mathbf{x}) \rVert}.
\end{equation}

\noindent
Here, ${D}_l^{des}$ is the distance map to the destination exit, and it is obtained as a solution to the continuous shortest path problem:

\begin{equation}\label{eq:shortestpath_a}
    \left\{ \begin{array}{ll}
        \lVert \nabla {D}_l^{des}(\mathbf{x}) \rVert = 1, & \textrm{if } \mathbf{x} \in \Omega \setminus \left\lbrace {\varepsilon}_l^{des} \cup W \right\rbrace,  \\
        {D}_{l}^{des}(\mathbf{x}) = 0, & \textrm{if } \mathbf{x} \in {\varepsilon}_l^{des}, \\
        {D}_l^{des}(\mathbf{x}) = \infty, & \textrm{if } \mathbf{x} \in W.
    \end{array}
    \right.
\end{equation}

\noindent
The problem is solved for all exits before the simulation is run. For detailed information on its numerical computation, see \citep{ourejor, fastmarchingmethod}.

If the passenger follows a guide, i.e., $\sum_g u_{lg} = 1$, it will instead move towards the guide's destination with the driving force,

\begin{equation}
    \mathbf{f}_{lg} = m_l \frac{v_l^0 \mathbf{e}_g (\mathbf{x}_l) - \mathbf{v}_l}{\tau}.
\end{equation}

\noindent
Here, the unit vector $\mathbf{e}_g$ is the direction towards the destination exit of guide $g$, ${\varepsilon}_g$. It is calculated in the same way as $\mathbf{e}_l^{des}$ in Eqs.~\eqref{eq:normedgradient} and~\eqref{eq:shortestpath_a}.

The term $\mathbf{f}_{li}$ includes the socio-psychological and physical forces between agents $l$ and $i \in I$, and similarly, the term $\mathbf{f}_{lw}$ contains the physical forces between agent $i$ and wall $w \subset W$. These have been extensively discussed in the article \citep{ourejor}; see its appendix for their exact mathematical expressions.

On the other hand, the change of velocity at time $t$ for guide $g$ is given by the equation of motion:

\begin{equation}\label{eq:eqofmotion_b}
    m_g \frac{d\mathbf{v}_g}{dt} = \mathbf{f}_{g}^0 + \sum_i \mathbf{f}_{gi} + \sum_{w \subset W} \mathbf{f}_{gw}.
\end{equation}

The term $\mathbf{f}_{gi}$ includes the socio-psychological and physical forces between guide $g$ and agent $i \in I$, and $\mathbf{f}_{lw}$ contains the physical forces between guide $g$ and wall $w \subset W$. These forces are of the same form as the ones for passengers. Finally, the term $\mathbf{f}_g^0$ is the driving force of guide $g$ to move towards its destination exit:

\begin{equation}
    \mathbf{f}_{g}^0 = m_l \frac{v_g^0 \mathbf{e}_g (\mathbf{x}_g) - \mathbf{v}_g}{\tau}.
\end{equation}

\section{Optimization framework}\label{sec:optimizationframework}

We assume that the crowd flow traffic patterns are uncertain. The uncertainty is related to the destination exits and desired speeds of the passengers, ${\varepsilon}_l^{des}$ and $v_l^0$, respectively. The possible realizations, or scenarios, of the uncertain parameters are denoted by ${\vartheta}^{k}, \ k \in \left\lbrace 1,...,K \right\rbrace$. In this section, we define the optimization variables and objective functions. The notation for risk measures is based on the article \citep{valueatrisk}.

\subsection{Probability definitions}\label{subsec:probdefs}

Let us denote the evacuation times of the individual agents to get out of the building with $t_i, \ i \in I=N \cup G $. The maximal element of these evacuation times equals the evacuation time of the crowd, which we denote by $T_{last}$. It is a random variable that depends on the set of scenarios $\theta = \left\lbrace {\vartheta}^{1},...,{\vartheta}^{K} \right\rbrace$. The evacuation time associated with scenario ${\vartheta}^{k}$ is $T_{last}^{k}$. Furthermore, each scenario is associated with a probability $p^k$, \ ${\sum}_{k} p^k = 1$.

We are dealing with a finite number of scenarios; hence Eq.~\eqref{eq:cdf}, the probability of $T_{last}$ being less than or equal to a number $\zeta \in \mathbb{R}$, is calculated from the discrete cumulative distribution function, by summing all the probabilities of the scenarios for which $T_{last}^{k}$ is less than $\zeta$,

\begin{align}\label{eq:cdf}
    \mathrm{P}\big(T_{last} \leq \zeta\big) = & \sum_{k, \ T_{last}^{k} \leq \zeta} \mathrm{P} \Big(T_{last}=T_{last}^{k}\Big)\nonumber \\
    = & \sum_{k, \ T_{last}^{k} \leq \zeta} p^k.
\end{align}

One of the objectives for our optimization problem is the mean of $T_{last}$, w.r.t., $\theta$. It is defined as a mean of a discrete random variable:

\begin{equation}\label{eq:mean}
    \mathbb{E}\left[T_{last}|\theta\right]=\sum_{k} p^k T_{last}^{k}.
\end{equation}

\noindent
From here on, we use the notation $\mathbb{E}\left[T_{last}\right]$ for $\mathbb{E}\left[T_{last}|\theta\right]$.

Next, we denote value-at-risk for probability level $\alpha \in (0,1)$ by $\mathrm{VaR}_{\alpha}$. For our problem, it is the smallest number $\zeta$ for which the cumulative probability $\mathrm{P}\big(T_{last} \leq \zeta\big)$ exceeds $\alpha$, and is defined as,

\begin{equation}\label{eq:var1}
    \mathrm{VaR}_{\alpha} \left[T_{last}\right] := \inf \left\lbrace \zeta \in \mathbb{R} : \mathrm{P}\big(T_{last} \leq \zeta\big)  \geq \alpha \right\rbrace.
\end{equation}

\noindent
We can rewrite Eq.~\eqref{eq:var1} by using Eq.~\eqref{eq:cdf}:

\begin{equation}\label{eq:var2}
    \mathrm{VaR}_{\alpha} \left[T_{last}\right] = \inf \Bigg\{ \zeta \in \mathbb{R} : \sum_{k, \ T_{last}^{k} \leq \zeta} p^k \geq \alpha \Bigg\} .
\end{equation}

Now, we can define our second objective, the conditional value-at-risk for probability level $\alpha$, $\mathrm{CVaR}_{\alpha}$. It is the conditional mean of $T_{last}$ exceeding $\mathrm{VaR}_{\alpha}$, and is defined as,

\begin{align}\label{eq:cvar1}
  \mathrm{CVaR}_{\alpha} \left[T_{last}\right] = & \mathbb{E} \left[T_{last} \big| T_{last} \geq \mathrm{VaR}_{\alpha} \left[T_{last}\right] \right].
\end{align}

\noindent
Again, because we have a finite number of scenarios, we can rewrite Eq.~\eqref{eq:cvar1} as a probability-weighted sum of those $T_{last}^{k}$ exceeding $\mathrm{VaR}_{\alpha}\left[T_{last}\right]$:

\begin{equation}\label{eq:cvar2}
    \mathrm{CVaR}_{\alpha} \left[T_{last}\right] = \mathrm{VaR}_{\alpha}\left[T_{last}\right]+ \frac{1}{1-\alpha} \sum_{k, \ T_{last}^{k} \geq \mathrm{VaR}_{\alpha}\left[T_{last}\right]} p^k \left[ T_{last}^{k} - \mathrm{VaR}_{\alpha}\left[T_{last}\right]\right].
\end{equation}

Note that in some situations, especially when there are a small number of scenarios, the worst-case scenario probability might equal or exceed the tail probability $1-\alpha$. Let us denote the worst-case scenario index by $\beta$, the worst-case scenario value, or the evacuation time associated with it, by $\mathrm{WCSV}$. Thus, it holds

\begin{equation}\label{eq:worstcase1}
    \beta = \arg \max_k \ T_{last}^k,
\end{equation}

and

\begin{equation}\label{eq:worstcase2}
    \mathrm{WCSV}\left[T_{last}\right] = T_{last}^{\beta}.
\end{equation}

When the worst-case scenario probability $p^{\beta} \geq 1-\alpha$, from Eq.~\eqref{eq:var2} we get,

\begin{equation}\label{eq:worstcase3}
    \mathrm{VaR}_{\alpha}\left[T_{last}\right] = \mathrm{WCSV}\left[T_{last}\right],
\end{equation}

\noindent
and furthermore, by inserting the above into Eq.~\eqref{eq:cvar2},

\begin{equation}\label{eq:worstcase4}
    \mathrm{CVaR}_{\alpha} \left[T_{last}\right] = \mathrm{WCSV}\left[T_{last}\right].
\end{equation}

So, in this situation, we can use the simpler $\mathrm{WCSV}$ instead of $\mathrm{CVaR}_{\alpha}$. In the example problem studied later in this paper, this will be the case.

\subsection{Optimization problem}

Next, we discretize the space $\Omega$ into square grid cells $\omega$, so that $\Omega \subset \cup \omega = \bar{\Omega}$. In our problem, the number $m$ of guides is fixed. Each guide $g \in G =\{n+1,...,n+m\}$ is associated with an origin grid cell $\omega_g  \subset \bar{\Omega}$, and a destination exit $\varepsilon_g \subset \mathcal{E}$. The optimization variables are feasible origin-destination pairs of the guides, i.e., feasible evacuation plans $\pi:=\left\{ (\omega_{n+1}, \varepsilon_{n+1}),...,(\omega_{n+m}, \varepsilon_{n+m}) \right\}, \ {\omega}_g \subset \bar{\Omega}, \ {\varepsilon}_g \subset \mathcal{E}, g \in G$. Let $\Pi$ be the set of all such evacuation plans.

The end-point conditions for our problem are the initial and final positions of the agents. The initial position of a guide $g$, $\mathbf{x}_g (0) \in \omega_g$, is a prespecified point in its corresponding origin grid cell $\omega_g$. The final position of a guide $g$, $\mathbf{x}_g (t_g) \in \varepsilon_g$, is any point in its corresponding destination exit $\varepsilon_g$. The initial positions of the agents are $\mathbf{x}_l(0)=\mathbf{x}_{l}^{0}, \ l \in N$. The agents can evacuate using any of the available exits; thus it holds for the final positions $\mathbf{x}_l (t_l) \in \mathcal{E}$.

We are interested in finding an evacuation plan that optimizes both the mean and $\mathrm{CVaR}_{\alpha}$ of $T_{last}$. Let us first define the following function:

\begin{equation}\label{eq:helpfunction}
  \phi(\pi, \theta) := T_{last}.
\end{equation}

\noindent
Then we can write our bi-objective optimization problem as

\begin{align}\label{eq:biobjective}
& \min_{\pi \in \Pi} \big\{\mathbb{E}\left[\phi(\pi,\theta)\right],\mathrm{CVaR}_{\alpha}\left[\phi(\pi,\theta)\right]\big\}; \nonumber \\
& \textrm{subject to Eqs.}~\eqref{eq:changeofvelocity},~\eqref{eq:eqofmotion_a},~\eqref{eq:eqofmotion_b}; \\
& \mathbf{x}_l (0)=\mathbf{x}_l^0, \ \mathbf{x}_l(t_l) \in \mathcal{E}, \ l \in N; \ \mathbf{x}_g(0) \in \omega_g, \ \mathbf{x}_g (t_g) \in \varepsilon_g, \ g \in G. \nonumber
\end{align}

\noindent
Here we are minimizing two objectives at the same time in the sense of Pareto-optimality. To compare two solutions, we use the concept of dominance. Solution ${\pi}^1$ dominates solution ${\pi}^2$ iff:

\begin{flalign}\label{eq:nondominated}
& \mathbb{E}\left[\phi({\pi}^1,\theta)\right] \leq \mathbb{E}\left[\phi({\pi}^2,\theta)\right], \ \textrm{and} \ \mathrm{CVaR}_{\alpha}\left[\phi({\pi}^1,\theta)\right] < \mathrm{CVaR}_{\alpha}\left[\phi({\pi}^2,\theta)\right],&\nonumber \\
\textrm{or} \quad \quad &&\\
& \mathbb{E}\left[\phi({\pi}^1,\theta)\right] < \mathbb{E}\left[\phi({\pi}^2,\theta)\right], \ \textrm{and} \ \mathrm{CVaR}_{\alpha}\left[\phi({\pi}^1,\theta)\right] \leq \mathrm{CVaR}_{\alpha}\left[\phi({\pi}^2,\theta)\right].& \nonumber
\end{flalign}

The set of solutions not dominated by any other solutions is called the set of nondominated, or Pareto-optimal, solutions. By definition, they are the solutions to the problem.

\section{Solution method}\label{sec:solutionmethod}

The bi-objective optimization problem of Eq.~\eqref{eq:biobjective} is solved with a combined numerical simulation and genetic algorithm (GA) procedure \citep{goldberg}. The GA searches iteratively for the nondominated solutions. At the same time, the evacuation simulation evaluates the objective function values, or fitnesses, of the found solutions and steers the randomized search process. The procedure ends when a convergence criterion is met. The procedure is summarized in the flowchart of Fig.~\ref{fig:algorithm_flowchart}, and it is explained next step-by-step.

\tikzstyle{decision} = [diamond, draw, fill=blue!20, 
    text width=5.75em, text badly centered, inner sep=0pt]
\tikzstyle{block} = [rectangle, draw, fill=blue!20, 
    text width=33em, minimum height=2em, text centered, inner sep=4pt]
\tikzstyle{line} = [draw, thick, -latex']
\tikzstyle{cloud} = [draw, ellipse,fill=red!20, minimum height=2em]

\begin{figure}[!ht]
    \centering
    \begin{tikzpicture}[node distance = 1.25cm, auto]
        \node [cloud] (start) {\footnotesize{Start}};
        \node [block, below of=start] (step1) {\footnotesize{ \textbf{Step 1.} Generate initial population randomly.}};
        \node [block, below of=step1] (step2) {\footnotesize{\textbf{Step 2.} Simulate evacuation plans under different scenarios.}};
        \node [block, below of=step2] (step3) {\footnotesize{\textbf{Step 3.} Calculate the bi-objective fitness value for the evacuation plans.}};
        \node [block, below of=step3, yshift=-0em] (step4) {\footnotesize{\textbf{Step 4.} Perform nondominated sorting (calculate nondomination rank and crowding distance).}};
        \node [block, below of=step4, yshift=-0em] (step5) {\footnotesize{\textbf{Step 5.} Apply selection, crossover, and mutation to create offspring population.}};
        \node [block, below of=step5, yshift=-0em] (step6) {\footnotesize{\textbf{Step 6.} Repeat Steps 2 and 3 for the offspring population.}};
        \node [block, below of=step6, yshift=-0em] (step7) {\footnotesize{\textbf{Step 7.} Repeat Step 4 for the union of the current generation and offspring populations.}};
        \node [block, below of=step7, yshift=-0em] (step8) {\footnotesize{\textbf{Step 8.} Choose the next generation population.}};
        \node [decision, below of=step8, yshift=-2.85em] (decide) {\footnotesize{Convergence criterion met?}};
        \node [block, below of=decide, yshift=-2.85em] (step9) {\footnotesize{\textbf{Step 9.} Return nondominated solutions.}};
        \node [cloud, below of=step9] (stop) {\footnotesize{Stop}};
        \coordinate [left of=step5, xshift=-14.85em] (A);
        \coordinate [left of=decide, xshift=-14.85em] (B);
    
        \path [line] (start) -- (step1);
        \path [line] (step1) -- (step2);
        \path [line] (step2) -- (step3);
        \path [line] (step3) -- (step4);
        \path [line] (step4) -- (step5);
        \path [line] (step5) -- (step6);
        \path [line] (step6) -- (step7);
        \path [line] (step7) -- (step8);
        \path [line] (step8) -- (decide);
        \path [line] (decide) -- node [left, near start] {\footnotesize{Yes}} (step9);
        \path [line] (step9) -- (stop);
        \draw [-, thick] (A) -- (B);
        \draw [-, thick] (decide) -| node [near start, above] {\footnotesize{No}} (B);
        \path [line] (A) -- (step5);
    \end{tikzpicture}
    \caption{Flowchart of the combined simulation and GA procedure.}
    \label{fig:algorithm_flowchart}
\end{figure}
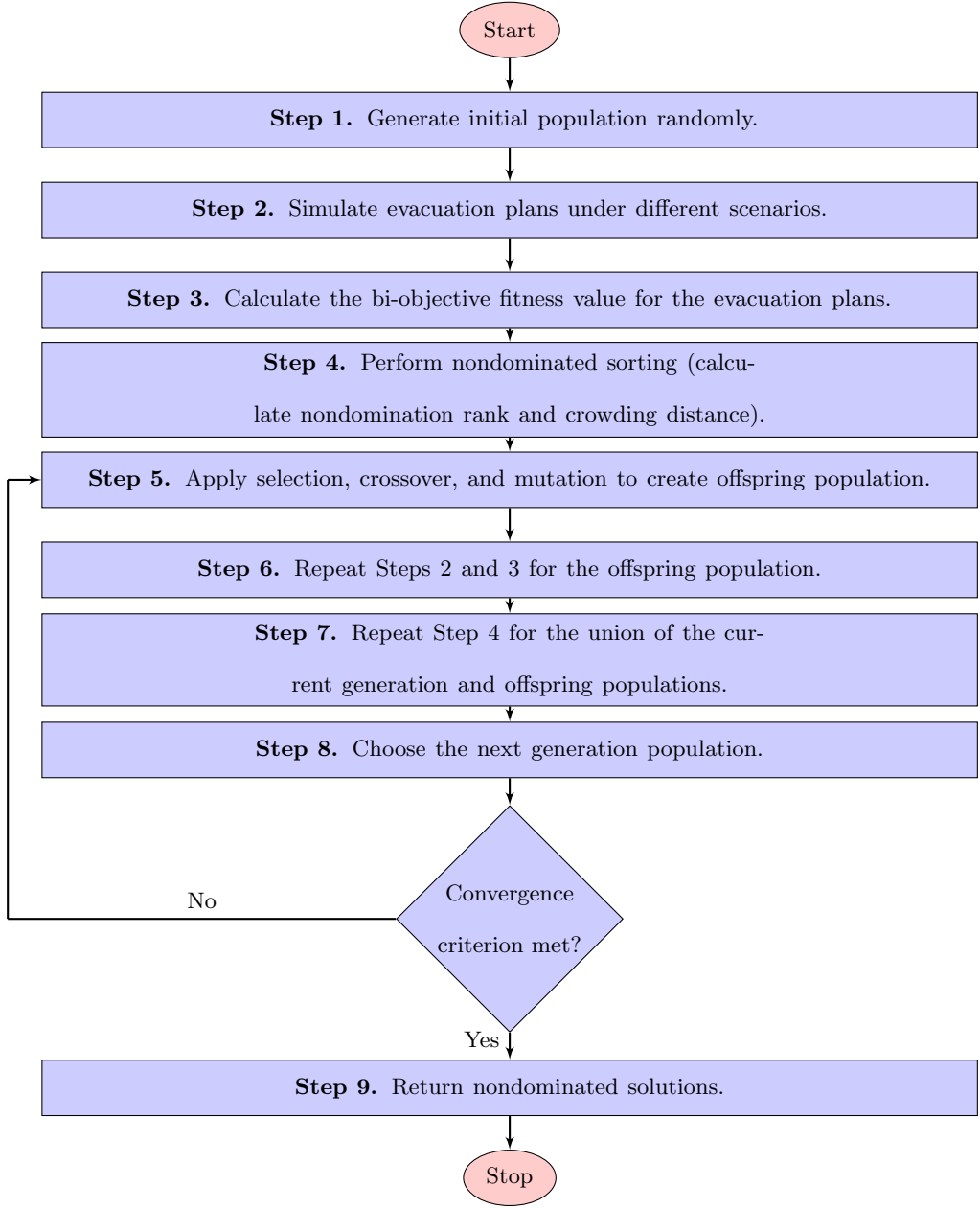

\textbf{Step 1.} In the first iteration, or generation, a random population with $Q$ number of solutions is created. The solutions are evacuation plans $\pi^1,...\pi^Q$. A gene in a solution contains the origin grid cell and destination exit of a single guide.

\textbf{Step 2.} The evacuation plans are simulated under different scenarios. More specifically, given an evacuation plan and a scenario ${\vartheta}^{k}$, the system defined by the constraints in Eq.~\eqref{eq:biobjective} is simulated with a numerical integration scheme to obtain $T_{last}^{k}$ (see, e.g., appendix of \citep{ownphysica} for further details). This is done for all evacuation plans in the population and all scenarios.

\textbf{Step 3.} The bi-objective fitness value is calculated with the equations from Sec.~\ref{subsec:probdefs}.

\textbf{Step 4.} The population solutions are ranked with the Nondominated Sorting Genetic Algorithm II (NSGA-II) \citep{deb2002fast}. The solutions are sorted based on dominance  (recall Eq.~\eqref{eq:nondominated}) to find the set, or front, of nondominated solutions. The first front gets the nondomination rank $0$. The first front's solutions are then set aside, and the sorting is repeated to find the next front. The solutions in the next front get the nondomination rank $1$. This procedure is repeated until all solutions have been ranked.

\begin{figure}
    \centering
    \begin{tikzpicture}[x=0.03cm,y=0.03cm]
        
        \draw[-latex, thick, draw=black] (10,60)--(300,60) node [below, xshift=-4cm] {\footnotesize{Mean (s)}};
        
        \draw[-latex, thick, draw=black] (10,60)--(10,300) node [above, rotate=90, xshift=-4cm] {\footnotesize{$\mathrm{CVaR}_{\alpha}$ (s)}};

        \coordinate (N1) at (50,250);
        \coordinate (N2) at (60,220);
        \coordinate (N3) at (115,150);
        \coordinate (N4) at (175,100);
        \coordinate (N5) at (220,80);
        
        \coordinate (C) at (60,100);
        
        \node [black] at (N1) {\LARGE{$\circ$}};
        \node [black] at (N2) {\LARGE{$\circ$}};
        \node [black] at (N3) {\LARGE{$\circ$}};
        \node [black] at (N4) {\LARGE{$\circ$}};
        \node [black] at (N5) {\LARGE{$\circ$}};
        \node [black] at ([shift=({-0.35cm,+0.35cm})]N2) {\footnotesize{${\pi}^{q-1}$}};
        \node [black] at ([shift=({-0.25cm,+0.25cm})]N3) {\footnotesize{${\pi}^q$}};
        \node [black] at ([shift=({-0.35cm,+0.35cm})]N4) {\footnotesize{${\pi}^{q+1}$}};

        \draw [color=black, ultra thick, dashed] ([shift=({+0.0cm,-0.25cm})]N2) -- (C) -- ([shift=({-0.25cm,+0.0cm})]N4);
        
        \node [black] at (90,260) {\LARGE{\textbullet}};
        \node [black] at (100,235) {\LARGE{\textbullet}};
        \node [black] at (155,175) {\LARGE{\textbullet}};
        \node [black] at (230,115) {\LARGE{\textbullet}};
        \node [black] at (250,90) {\LARGE{\textbullet}};
        \node [black] at (250,105) {\LARGE{\textbullet}};
        \node [black] at (260,150) {\LARGE{\textbullet}};
        
        \coordinate (BB1) at (155,290);
        \coordinate (BB2) at (155,250);
        \coordinate (BB3) at (295,250);
        \coordinate (BB4) at (295,290);
        \draw [color=black] (BB1) -- (BB2) -- (BB3) -- (BB4) -- cycle;
        \node [black] at (165,280) {\LARGE{$\circ$}};
        \node [black] at (165,260) {\LARGE{\textbullet}};
        \node [black] at (235,281) {\footnotesize{Nondominated solutions}};
        \node [black] at (225,261) {\footnotesize{Dominated solutions}};
\end{tikzpicture}
    \caption{The crowding distance of solution ${\pi}^q$. It is the $L^1$ distance between its neighboring solutions ${\pi}^{q-1}$ and ${\pi}^{q+1}$ (the sum of the lengths of the two dashed lines). The horizontal axis gives the solutions' mean evacuation time, whereas the vertical axis gives the $\text{CVaR}_{\alpha}$ evacuation time.}
    \label{fig:crowdingdistance}
\end{figure}
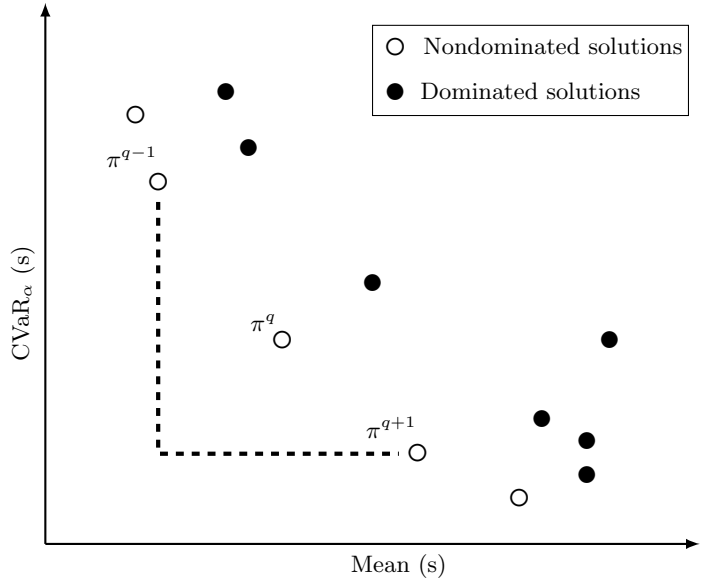

After nondomination ranks have been assigned, a crowding distance is calculated for each solution \citep{fortin2013revisiting}. It estimates the density of solutions with the same rank surrounding a particular solution. For a given solution, it calculates the $L^1$ distance between its two neighboring solutions in the objective space (see Fig.~\ref{fig:crowdingdistance}). Solutions with a higher crowding distance, i.e., are located in a less crowded region, are preferred. This is because solutions located close to each other, in the sense of objective functions' values, are likely to have evacuation plans close to each other. There should be diversity in the solutions that the GA does not get stuck in a local optimum.

\textbf{Step 5.} The offspring solutions are obtained by applying selection, crossover, and mutation operations on the current generation, or parent, solutions. In selection, we use the unique fitness tournament selection \citep{fortin2013revisiting}. In it, a number of solutions with unique fitness values are sampled. Each sampled solution is paired against one other sampled solution. In the paired contest, the solution with a lower nondomination rank wins and is selected to undergo further operations. If the solutions have the same rank, the one with a higher crowding distance wins the paired contest. This process is repeated until $Q$ solutions have been selected.

After selection, each selected solution is paired with one other selected solution for crossover. We use the single-point crossover operation, where the two parent solutions create offspring solutions that contain half of the genetic material of both parents. Thus, the offspring solutions contain half of the guides of each parent's evacuation plan. The crossover operator is applied with a certain probability. If it is not applied, the offspring solutions have the same genetic information as to their parents.

Finally, a mutation operator is applied to the offspring solutions. It is applied separately, with a certain probability on each gene. It can either change the origin grid cell of a guide, its destination exit, or both of them.

\textbf{Step 6.} The bi-objective fitness of the offspring solutions is calculated by repeating Steps 2 and 3.

\textbf{Step 7.} Step 4 is repeated for the union of the current generation and offspring populations.

\textbf{Step 8.} For the next-generation population, $Q$ solutions are chosen from the current generation and offspring populations' union. The solutions with the lowest nondomination ranks are chosen. If all the same rank solutions do not fit, those with the highest crowding distance are prioritized.

\textbf{Step 9.} Return the first nondominated front calculated in Step 7.

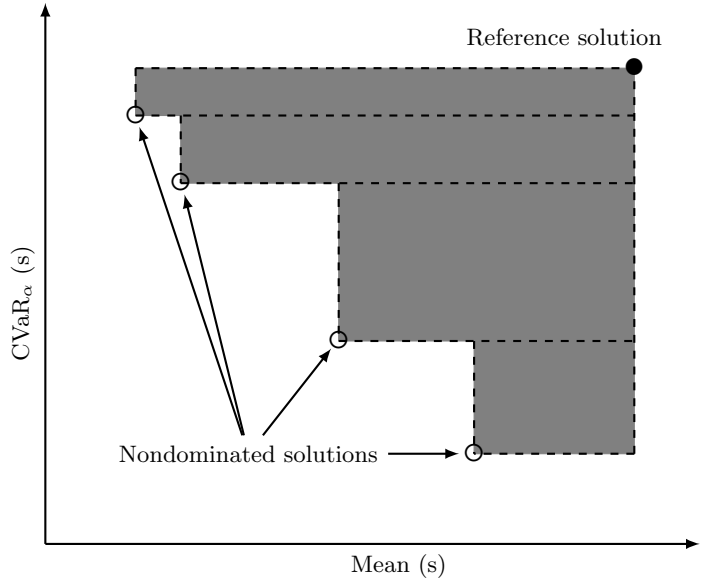
\begin{figure}
    \centering
    \begin{tikzpicture}[x=0.03cm,y=0.03cm]
        
        \coordinate (nadir) at (271,271);
        \coordinate (H1) at (50,271);
        \coordinate (N1) at (50,250);
        \coordinate (H2) at (70,250);
        \coordinate (N2) at (70,220);
        \coordinate (H3) at (140,220);
        \coordinate (N3) at (140,150);
        \coordinate (H4) at (200,150);
        \coordinate (N4) at (200,100);
        \coordinate (H5) at (271,100);
        
        \coordinate (L1) at (271,150);
        \coordinate (L2) at (271,220);
        \coordinate (L3) at (271,250);
    
        \draw[-latex, thick, draw=black] (10,60)--(300,60) node [below, xshift=-4cm] {\footnotesize{Mean (s)}};
        
        \draw[-latex, thick, draw=black] (10,60)--(10,300) node [above, rotate=90, xshift=-4cm] {\footnotesize{$\mathrm{CVaR}_{\alpha}$ (s)}};

        \node [black] (nondtxt) at (100,100) {\footnotesize{Nondominated solutions}};
        \draw[-latex, thick] (nondtxt) -- ([shift=({+0.05cm,-0.15cm})]N1);
        \draw[-latex, thick] (nondtxt) -- ([shift=({+0.05cm,-0.15cm})]N2);
        \draw[-latex, thick] (nondtxt) -- ([shift=({-0.1cm,-0.1cm})]N3);
        \draw[-latex, thick] (nondtxt) -- ([shift=({-0.2cm,-0.0cm})]N4);

        
        \filldraw [draw=gray, fill=gray] (H1) -- (nadir) -- (L3) -- (N1) -- cycle;
        
        \filldraw [draw=gray, fill=gray] (H2) -- (L3) -- (L2) -- (N2) -- cycle;
        
        \filldraw [draw=gray, fill=gray] (H3) -- (L2) -- (L1) -- (N3) -- cycle;
        
        \filldraw [draw=gray, fill=gray] (H4) -- (L1) -- (H5) -- (N4) -- cycle;
        
        \path[-]
        (H1) edge[black, dashed, thick] (nadir)
             edge[black, dashed, thick] (N1)
        (N1) edge[black, dashed, thick] (L3)
        (H2) edge[black, dashed, thick] (N2)
        (N2) edge[black, dashed, thick] (L2)
        (H3) edge[black, dashed, thick] (N3)
        (N3) edge[black, dashed, thick] (L1)
        (H4) edge[black, dashed, thick] (N4)
        (H5) edge[black, dashed, thick] (N4)
             edge[black, dashed, thick] (nadir);

        \node [black] at (N1) {\LARGE{$\circ$}};
        \node [black] at (N2) {\LARGE{$\circ$}};
        \node [black] at (N3) {\LARGE{$\circ$}};
        \node [black] at (N4) {\LARGE{$\circ$}};
        
        \node [black] at (nadir) {\LARGE{\textbullet}};
        \node [black] at (240,285) {\footnotesize{Reference solution}};

\end{tikzpicture}
    \caption{Hypervolume of a set of nondominated solutions. It is the sum of the rectangular gray areas defined by the nondominated solutions and the reference solution. The horizontal axis gives the solutions' mean evacuation time, whereas the vertical axis gives the $\text{CVaR}_{\alpha}$ evacuation time.}
    \label{fig:hypervolume}
\end{figure}

After Step 8, we check if a convergence criterion is met. More precisely, to measure convergence, we use a hypervolume indicator \citep{hypervolume}. To calculate its value, we first construct a reference solution. The reference solution is a point in the objective space that has larger $\mathrm{CVaR}_{\alpha}$ and mean evacuation time than any feasible solution. The hypervolume indicator is calculated as the sum of the rectangular areas defined by the reference solution and the first nondominated front from Step 7 (see Fig.~\ref{fig:hypervolume}).

The hypervolume indicator is a measure of the quality of the set of nondominated solutions. The solutions of a bi-objective optimization problem give the largest hypervolume. When the hypervolume indicator has not increased for a predefined number of generations, we consider the algorithm to have converged. Note that the choice of the reference solution can affect the convergence of NSGA-II.

\section{Case study: evacuation of a passenger terminal}
\label{sec:problemsetting}

Here, we have in mind a busy passenger interchange terminal. Passengers with different characteristics are arriving there through one exit and later departing from another. At such places, common walking areas can get very crowded at times \citep{schultzandfricke, ali2019impact}. The changing crowd conditions make planning an evacuation particularly difficult. The question we ask is: how should the evacuation planner instruct the safety staff, or guides, to lead the passengers out of the terminal in case of emergency when accounting for different scenarios?

More specifically, if the passengers have just arrived, they move to the exits on the opposite end of a hallway. If they are departing, they move to their nearest exits. Passengers move in groups either with regular or slow speed. Slow-moving passengers are elderly or somehow disabled. Note that crowds usually are heterogeneous. We assume that the passengers move at the same speed for demonstration purposes.

\begin{figure}[!h]
  \centering
  \subfigure[Scenarios 2 and 3]{\includegraphics[width=0.49\textwidth]{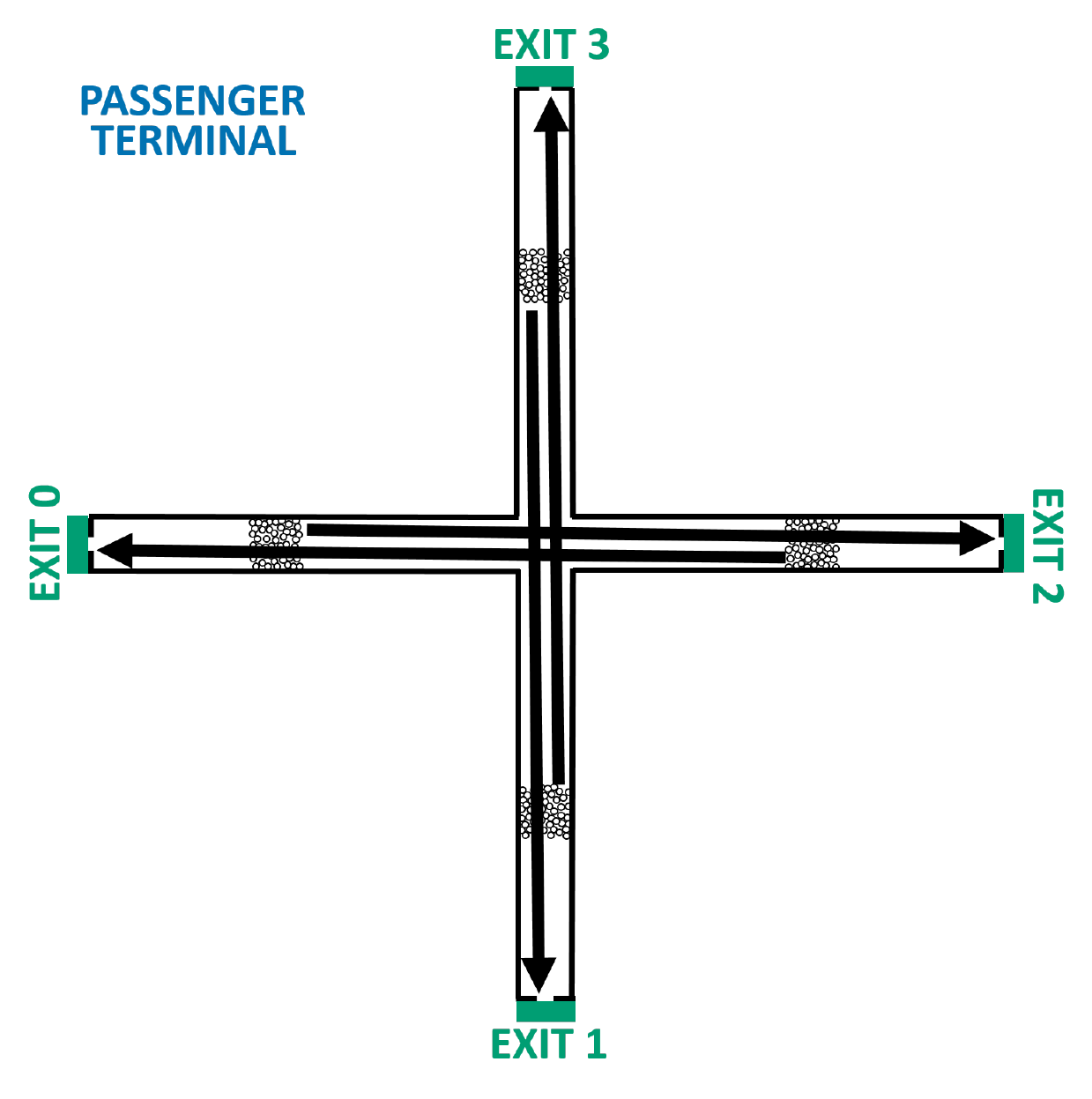}}
   \subfigure[Scenarios 1 and 4]{\includegraphics[width=0.49\textwidth]{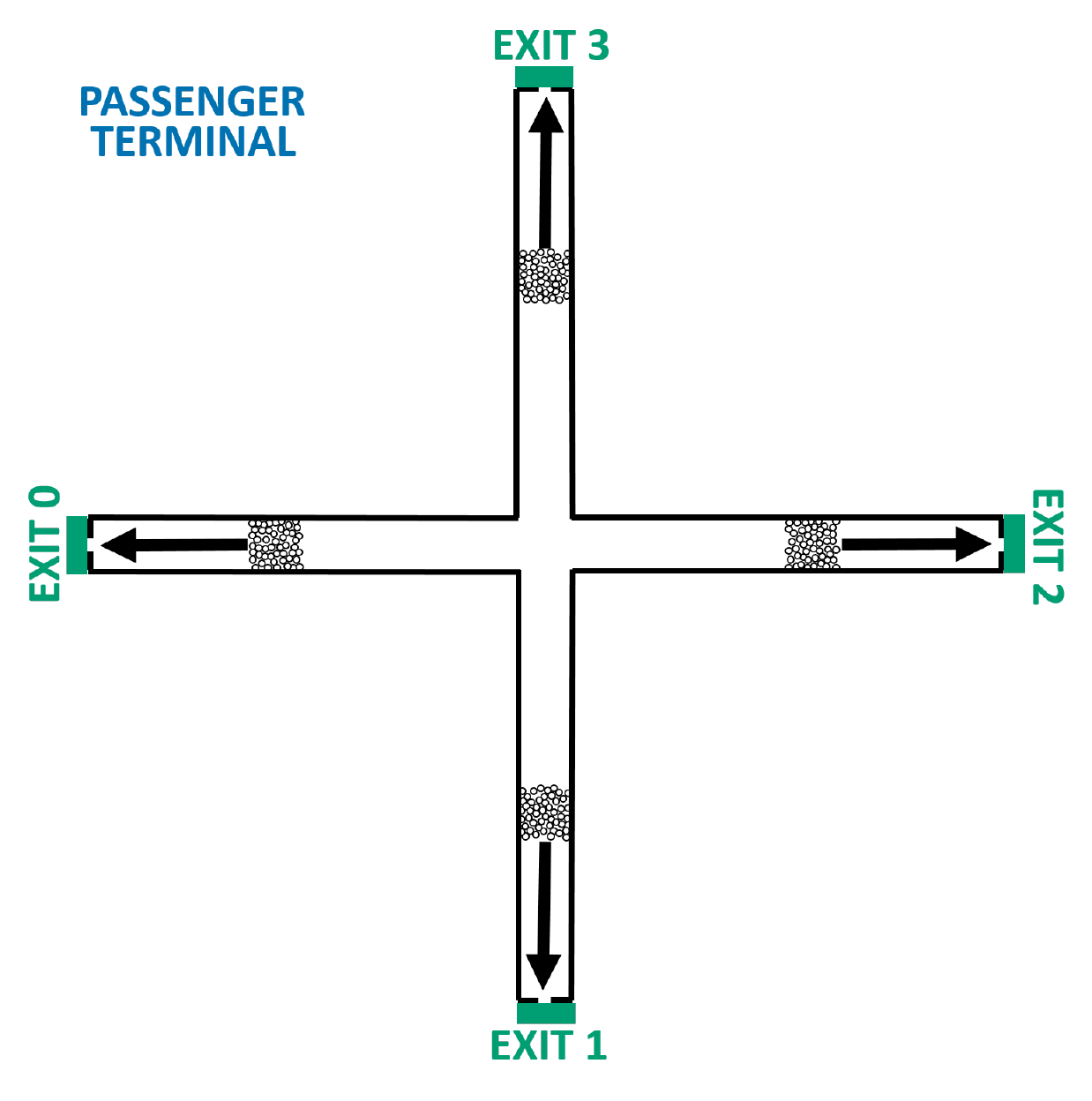}}
  \caption{A schematic diagram of possible scenarios in a passenger terminal: (a) arriving agents in Scenario 2 (slow) and Scenario 3 (fast), (b) departing agents in Scenario 1 (slow) and Scenario 4 (fast). The terminal halls are $5$ m wide, $85$ m long, and the distance from the exit to the intersection area is $40$ m. The exit widths are $1.2$ m.}
  \label{fig:departure}
\end{figure}

We create four characteristic scenarios, which are depicted in the schematic diagram of the fictional interchange terminal in Fig.~\ref{fig:departure}. We assign the probabilities $0.3$, $0.2$, $0.2$ and $0.3$ to the scenarios, respectively. Also, we set the probability level $\alpha=0.95$ for the risk measure. Thus, in this example, minimizing ${\textrm{CVaR}}_{\alpha}$ equals to minimizing the evacuation time of the worst-case scenario.

In Scenarios 2 and 3 agents are arriving (see Fig.~\ref{fig:departure}(a)), while in Scenarios 1 and 4 the agents are departing (see Fig.~\ref{fig:departure}(b)). One can think of Scenarios 2 and 3 representing the same agents' movement as in the other two scenarios, but earlier, when they have just arrived in the terminal. In Scenarios 1 and 2, the agents move slowly, while in Scenarios 3 and 4, they move fast.

\begin{figure}[!h]
\centering
\includegraphics[width=0.5\textwidth]{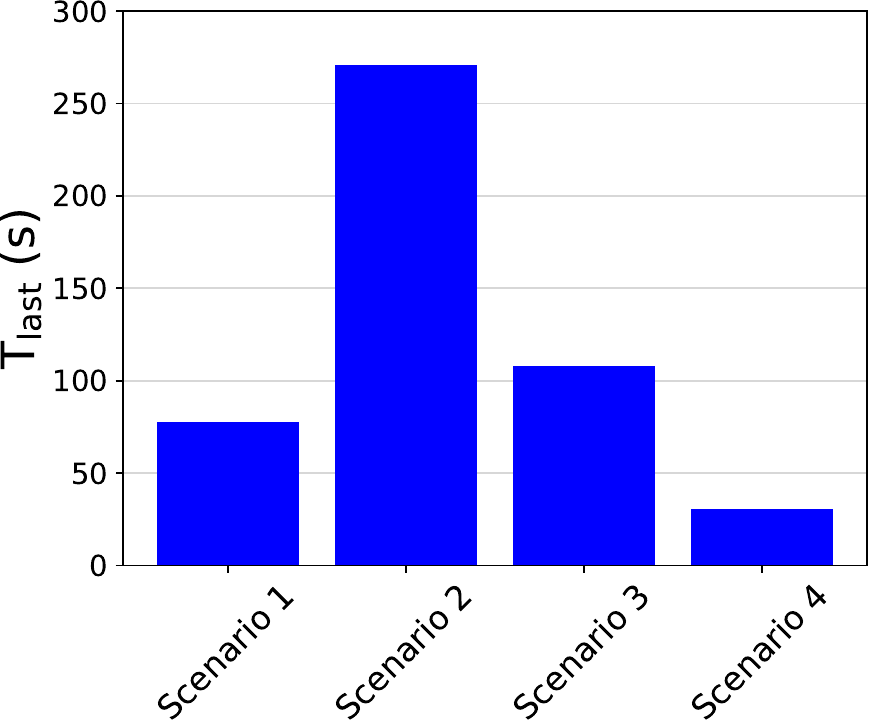}
  \caption{Evacuation time $T_{last}$ for different scenarios. A large congestion in the worst-case scenario (Scenario 2) results in a slow evacuation.}
  \label{fig:scenarios}
\end{figure}

In an emergency, like a bomb threat, we assume that passengers will go about their business unless guided otherwise. Thus, in an unguided evacuation, the evacuation time of the crowd equals the time for them to reach their destination exits (see Fig.~\ref{fig:scenarios}). The evacuation times of Scenarios 2 and 3 are the longest since the arriving agents create a four-way counterflow at the intersection, which quickly causes a large congestion. The agents also have a longer walk when they move to the opposite end of the hallway. Moreover, in Scenario 2, the agents move slowly, which means that it is noticeably the slowest of all four scenarios.

Here our focus is only on how to use guides in an evacuation. We acknowledge that in reality, for example, loudspeakers could also give information to the passengers.  We are interested in studying the tradeoff between the number of guides used and the crowd's evacuation time. Furthermore, we assume guides are instructed for one evacuation plan by the evacuation planner, and they will use it for all scenarios.

Initially, each passenger agent group is located in the middle of one leg of the terminal. In each of the four groups there are $50$ agents. For the circle-shaped passenger agents $l \in N$, the initial positions $\mathbf{x}_l^0$, radii $r_l$ and masses $m_l$ are fixed for all scenarios. Before fixing the values, the parameters $m_l$ and $r_l$ are drawn from a truncated normal distribution with a cutoff at three times the standard deviation. The mean and standard deviations are $73.5$ kg and $8.0$ kg, and $0.255$ m and $0.035$ m, respectively for $m_l$ and $r_l$. The reaction time is set $\tau=0.5$ s for all agents. These values are all taken from the FDS+Evac user manual \citep{korhonen2009fire}. In Scenarios 1 and 2, the desired speeds of passengers are $v_l^0=0.5$ m/s, whereas for Scenarios 3 and 4 they are $v_l^0=1.55$ m/s. Also, as stated above, the destination exits ${\varepsilon}_l^{des}$ of passengers vary between scenarios. On the other hand, for the circle-shaped guide agents $g \in G$, we set the values of a typical male: $m_g=80$ kg, $r_g=0.27$ m, and $v_g^0=1.15$ m/s. Also, the interaction range of guide agents is set to $r_{guide}=10$ m. The origin grid cells for guides, $\omega \in \bar{\Omega}$, are $2 \textrm{ m} \times 2 \textrm{ m}$.

\section{Numerical results}\label{sec:numericalresults}

\subsection{Implementation details and performance}

We solve the bi-objective optimization problem for the fictional passenger terminal for $1$, $2$, $3$ and $4$ guides, with the procedure presented in Fig.~\ref{fig:algorithm_flowchart}. Typically, to solve an optimization problem with a GA, the algorithm parameters are tuned manually in a problem-specific manner. Here, we extensively tried different algorithm parameters to assure the most accurate and efficient convergence. We set the GA population size, i.e., number of solutions to be evaluated in each generation, or iteration, to $Q=40$, crossover probability to $0.85$ and mutation probability to $0.10$. We consider the procedure to have converged when the hypervolume has not increased for $15$ consecutive generations. When calculating the hypervolume (recall Fig.~\ref{fig:hypervolume}), we set $271$ s as the objective function value for the reference solution for both ${\textrm{CVaR}}_{\alpha}$ and mean evacuation time. The reason being that, in the unguided evacuation, in Scenario $2$, the evacuation time is $271$ s, and we assume the solutions cannot have worse objective function values than that.

The crowd simulation model is implemented in Python code. Some of the code's core parts are written as Numba-decorated functions, which translates Python functions to optimized machine code at runtime. Numba-compiled numerical algorithms in Python can approach the speeds of C or FORTRAN \citep{numba}. The GA is implemented in Bash script that calls the crowd simulation code written in Python. For reproducibility, all codes are published \citep{articlecode}. The procedure has been run on the Aalto University high-performance computing cluster Triton. A single generation of the GA has been run parallel on Triton using its computing nodes that are Intel Xeon X5650 $2.67$ GHz with $48$ GB or $96$ GB memory, and Xeon E5 2680 v2 $2.80$ GHz with $64$ GB or $256$ GB memory.

Simulation of one generation takes a maximum of $10$ min. The algorithm converges in $27$, $34$, $58$ and $32$ generations for $1$, $2$, $3$ and $4$ guides, respectively. In computation time, this is approximately $4$ h $30$ min, $5$ h $40$ min, $9$ h $40$ min and $5$ h $20$ min. It is not surprising that when we increase the amount of guides, i.e., optimization variables, the computation time increases. However, with $4$ guides, the computation time decreases again. This could be related to the fact that there is only one Pareto-optimal solution with $4$ guides, as we will see soon.

\subsection{Pareto-optimal evacuation plans}

\begin{figure}[!h]
  \centering
  \includegraphics[width=0.5\textwidth]{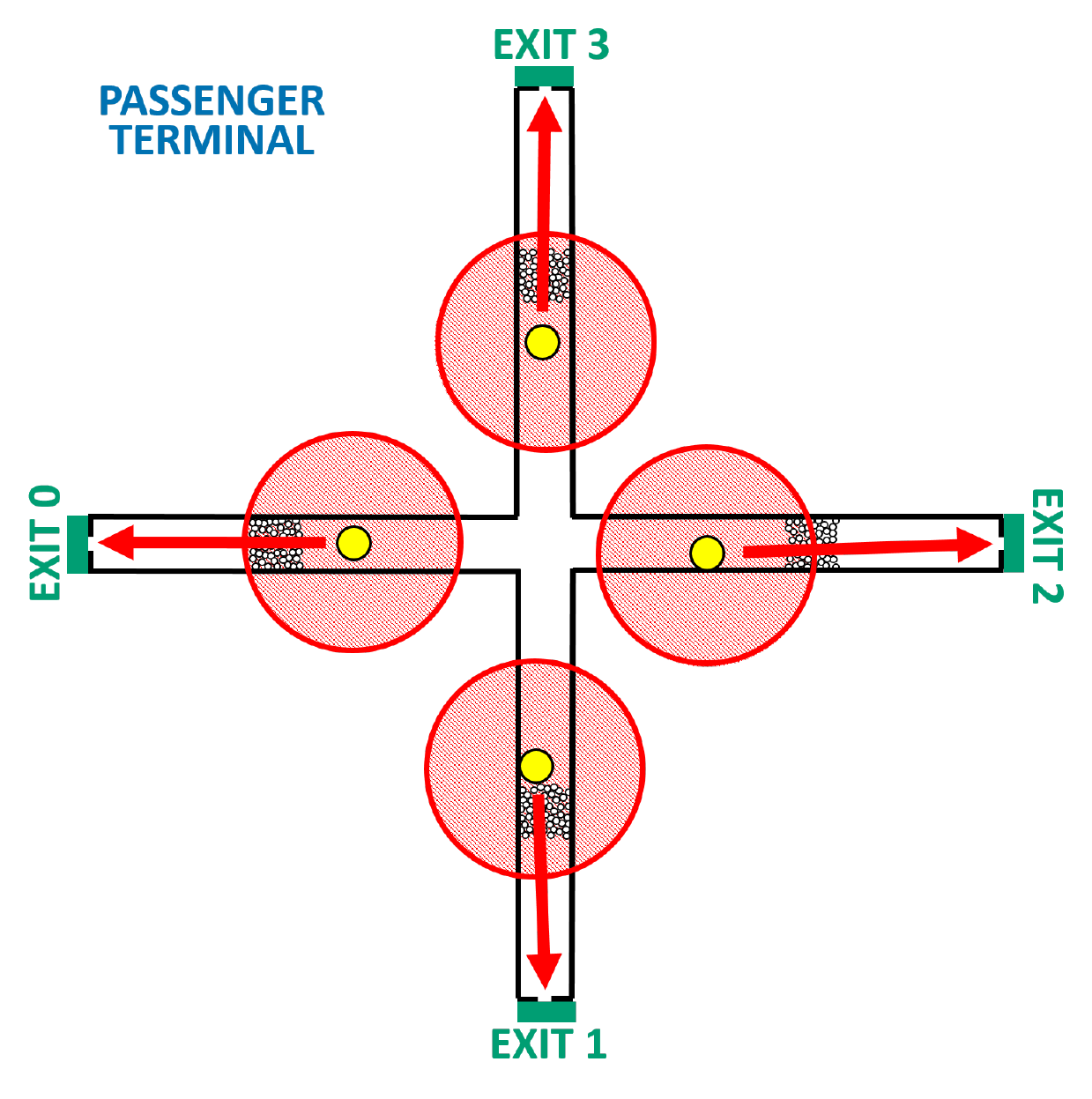}
  \caption{The solution that minimizes both ${\textrm{CVaR}}_{\alpha}$ and mean evacuation time is obtained with $4$ guides. The red circles represent the guides' interaction range, and the red arrows their paths to their destination exits.}
  \label{fig:4guides}%
\end{figure}

Note that the GA is a heuristic rather than an exact solution algorithm. Thus it is appropriate to call the solutions near-optimal instead of optimal. However, for readability, we drop the prefix \textit{near-}. We can deduce from the problem setting that if there is no restriction on the number of guides, the solution would be to take the agent groups to their nearest exits, which can be done using $4$ guides. It results in minimum evacuation time for all scenarios. The algorithm finds the $4$ guide optimum; see Fig.~\ref{fig:4guides}. So, we are confident that the other obtained solutions are also optimal. 

\begin{figure}[!h]
  \centering
  \includegraphics[width=0.6\textwidth]{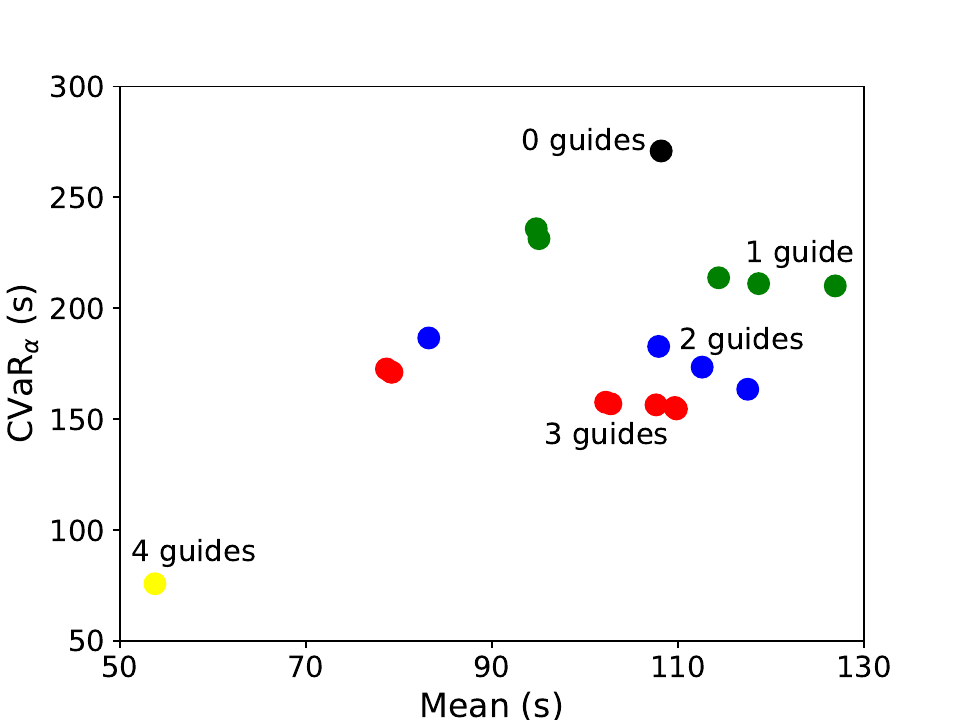}
  \caption{Objective function values for the found Pareto-optimal solutions. The same-colored circles depict the solutions with the same number of guides. The horizontal axis gives the solutions' mean evacuation time,  whereas the vertical axis gives the $\text{CVaR}_{\alpha}$ evacuation time.}
  \label{fig:pareto_ev_cvar}
\end{figure}

The found Pareto-optimal solutions, or Pareto fronts, for the different number of guides, are all presented in Fig.~\ref{fig:pareto_ev_cvar}. For $1$, $2$, and $3$ guides, there are multiple solutions, which means that there is a tradeoff between minimizing ${\textrm{CVaR}}_{\alpha}$ and mean evacuation time. On each Pareto front, the solution down on the right minimizes ${\textrm{CVaR}}_{\alpha}$, and is called ${\textrm{CVaR}}_{\alpha}$-optimal solution. Whereas the solution up on the left minimizes the mean evacuation time and is thus called the mean-optimal solution. We can see that the GA only finds a few solutions per Pareto front. There exist more solutions, and if the algorithm were able to find them, the fronts would look denser. By tuning algorithm parameters, we can obtain more solutions. However, we do not find it necessary, as we will see below, the ${\textrm{CVaR}}_{\alpha}$- and mean-optimal solutions already characterize all solutions on a Pareto front.

\begin{figure}[ht!]
\centering
\subfigure[]{\includegraphics[width=0.49\textwidth]{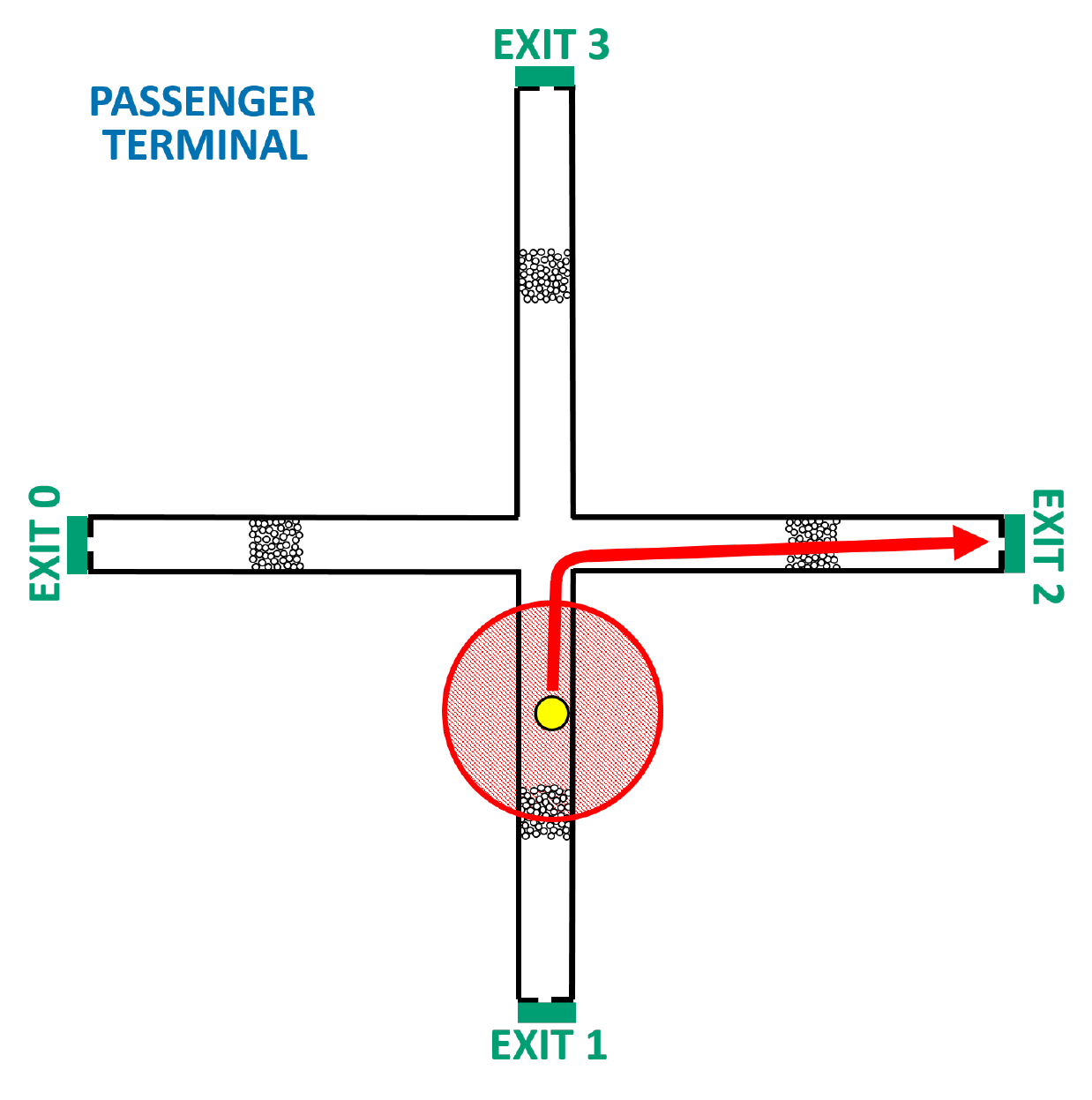}}
\subfigure[]{\includegraphics[width=0.49\textwidth]{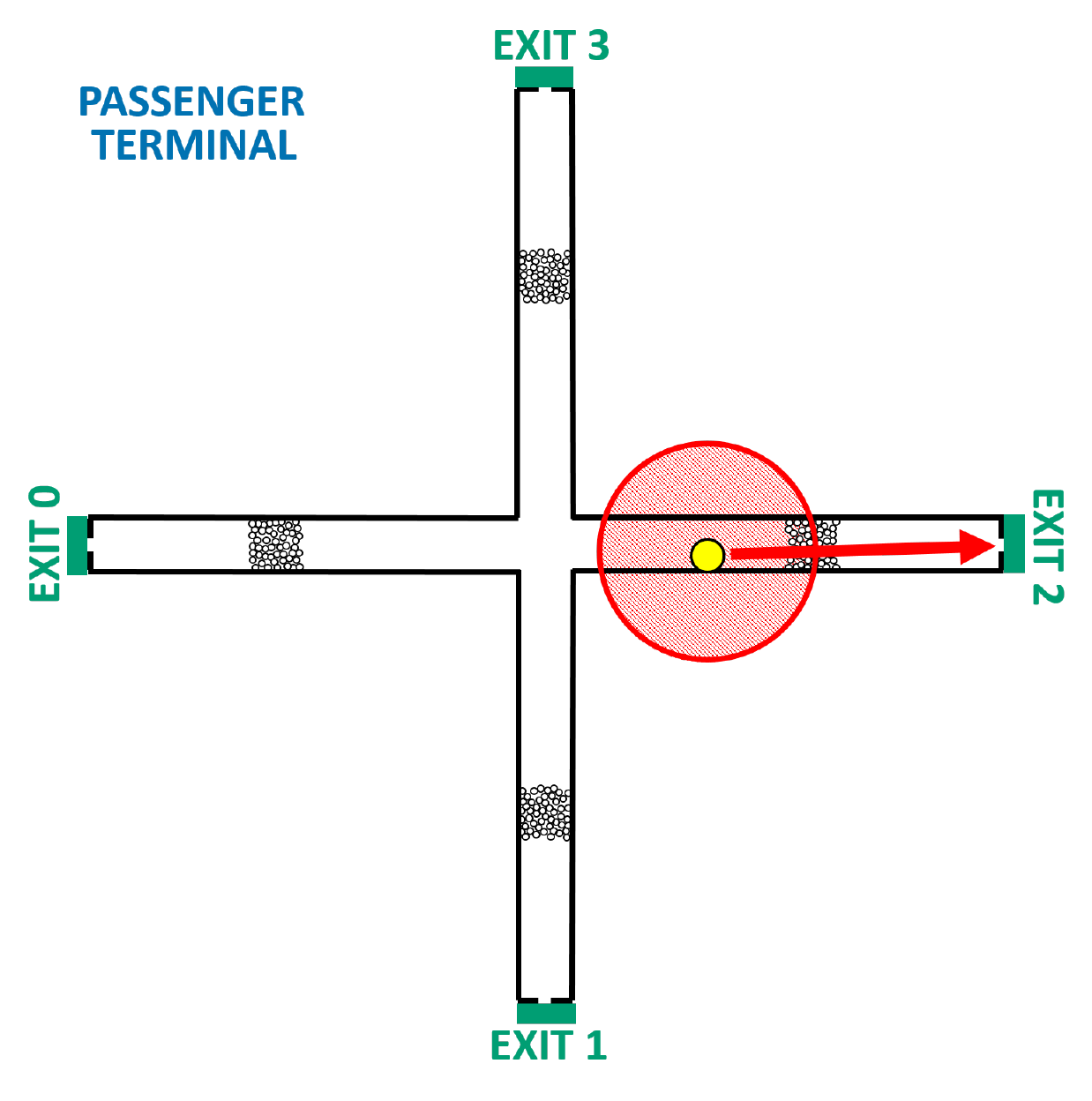}}
\caption{(a) ${\textrm{CVaR}}_{\alpha}$-, and (b) mean-optimal evacuation plan with $1$ guide.}
\label{fig:1guide}
\end{figure}

\begin{figure}[ht!]
\centering
\subfigure[]{\includegraphics[width=0.49\textwidth]{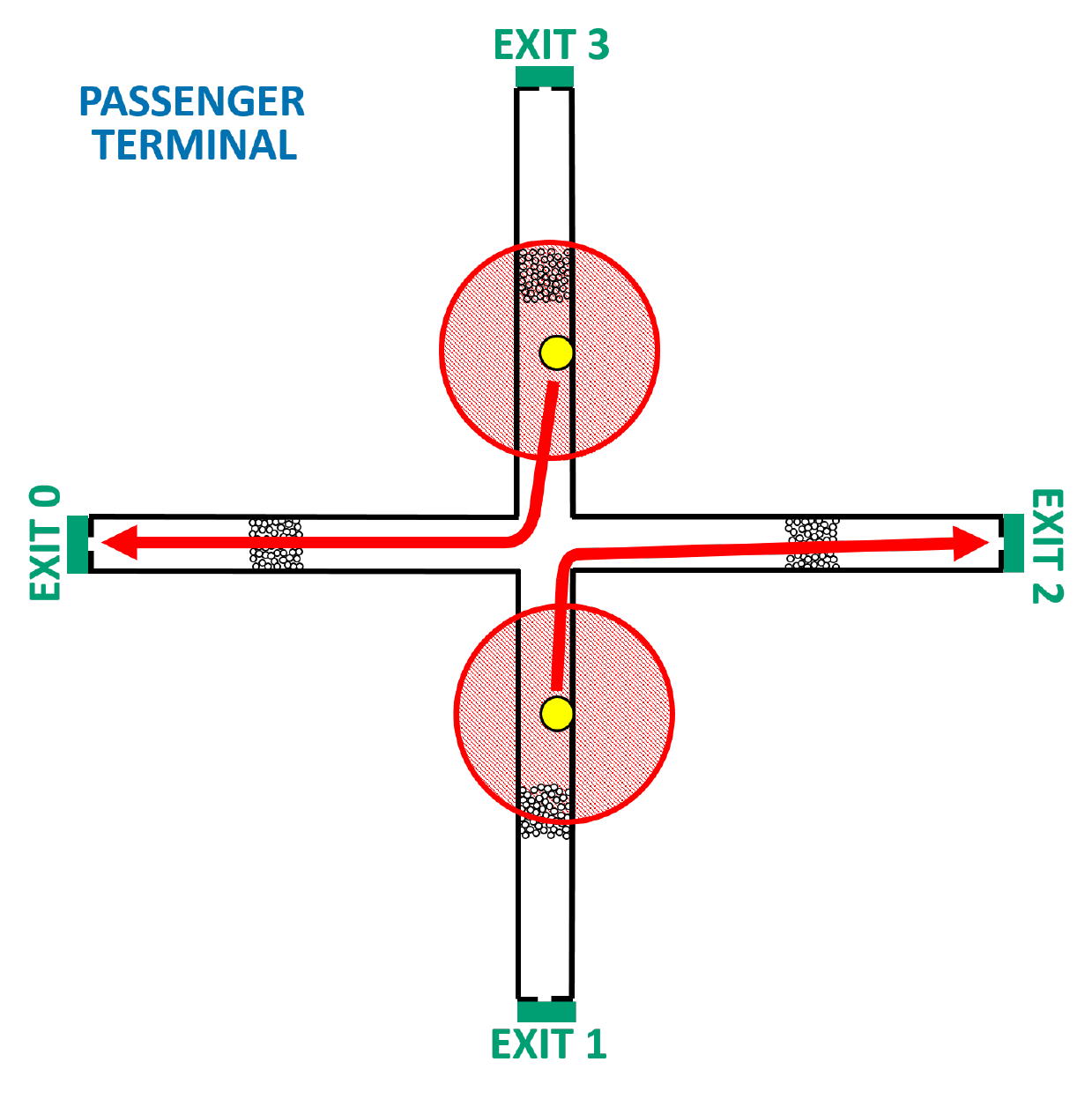}}
\subfigure[]{\includegraphics[width=0.49\textwidth]{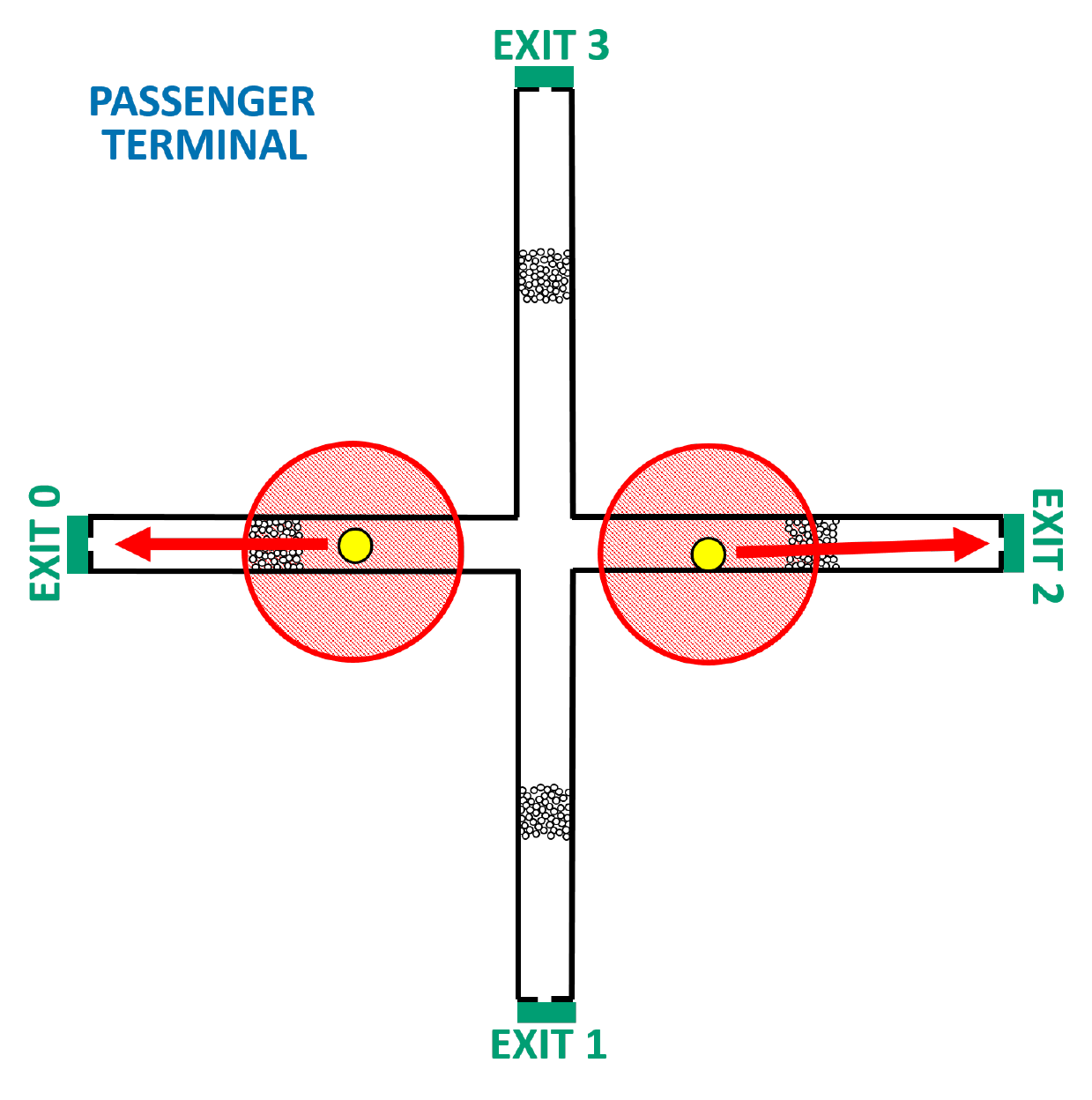}}
\caption{(a) ${\textrm{CVaR}}_{\alpha}$-, and (b) mean-optimal evacuation plan with $2$ guides.}
\label{fig:2guides}
\end{figure}

\begin{figure}[ht!]
\centering
\subfigure[]{\includegraphics[width=0.49\textwidth]{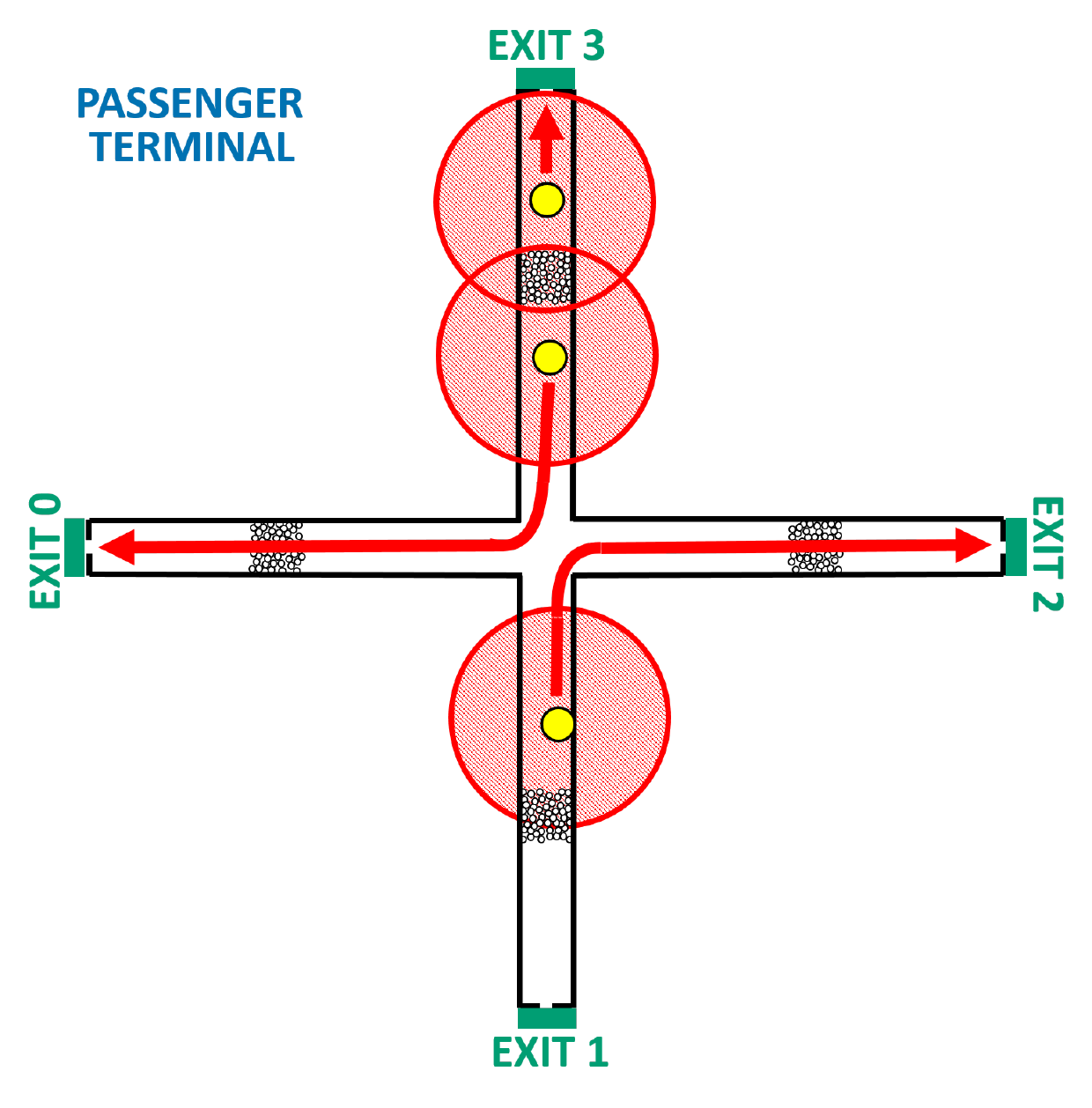}}
\subfigure[]{\includegraphics[width=0.49\textwidth]{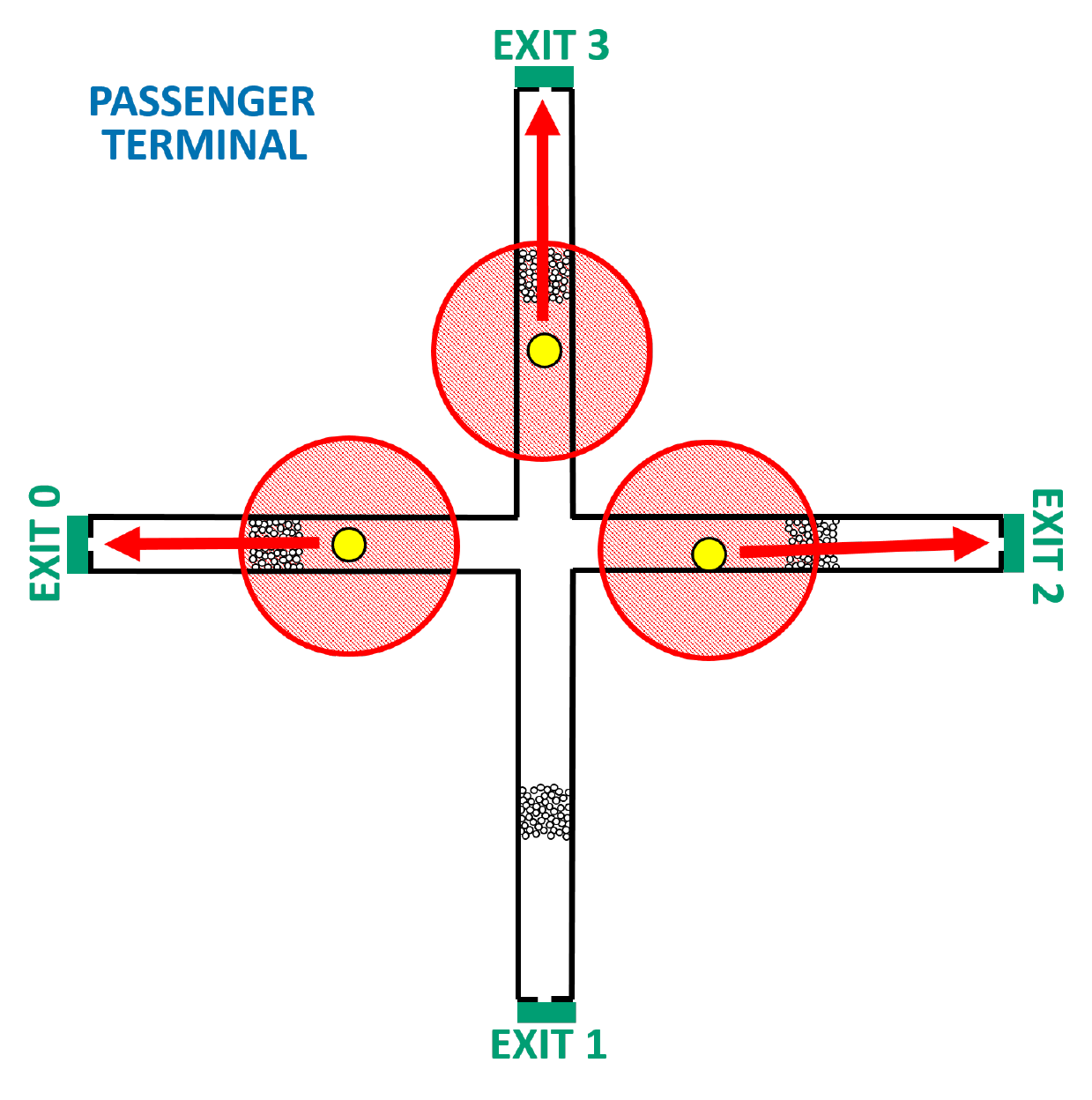}}
\caption{(a) ${\textrm{CVaR}}_{\alpha}$-, and (b) mean-optimal evacuation plan with $3$ guides.}
\label{fig:3guides}
\end{figure}

Let us study the optimal evacuation plans for $1, 2$, and $3$ guides more closely. Figs.~\ref{fig:1guide}, ~\ref{fig:2guides}, ~\ref{fig:3guides} present schematic diagrams of them. In each figure, the left subfigure represents the ${\textrm{CVaR}}_{\alpha}$-optimal evacuation plan, while the right figure represents the mean-optimal evacuation plan. The figures depict the initial situation when using these plans. The red circles describe the guide's interaction range, and the red arrow describes the path the guide takes to its destination exit. Along the path, the guide interacts with agents within the interaction range, and these agents are led to the guide's destination exit.

Note that the exit assignments are the same both in the ${\textrm{CVaR}}_{\alpha}$- and mean-optimal solutions (see Figs.~\ref{fig:1guide},~\ref{fig:2guides},~\ref{fig:3guides}). However, initially, in the mean-optimal solutions, the guides are closer to their assigned exits. As shown above, in Fig.~\ref{fig:pareto_ev_cvar}, there are multiple Pareto-optimal solutions for a fixed number of guides. When we monitored the solutions more closely, we noticed that moving on the Pareto front from the ${\textrm{CVaR}}_{\alpha}$- to the mean-optimal solution, the guides' initial positions are set closer to their destination exit. The only exception is the mean-optimal solution in Fig.~\ref{fig:3guides}(b), where the upper guide is positioned a little farther away in the mean-optimal solution. This does not matter since it still influences the upper group of agents.

If we think about the unguided situation, the slowest evacuation occurs in Scenario $2$. In it, agents arrive and head to the exit gates on the opposite end of the hallway, resulting in a four-way counterflow and large congestion at the intersection. Hence, the main feature of ${\textrm{CVaR}}_{\alpha}$-optimal solutions is to decrease congestion effects at the intersection. On the other hand, in the most probable Scenarios $1$ and $4$, the agents head to the nearest exits even without guides. Thus, in the mean-optimal solutions, guides are set to lead the agents to the nearest exits.

\subsection{Congestion at the intersection}

\begin{figure}[htb!]
\centering
\subfigure[]{\includegraphics[width=0.49\textwidth]{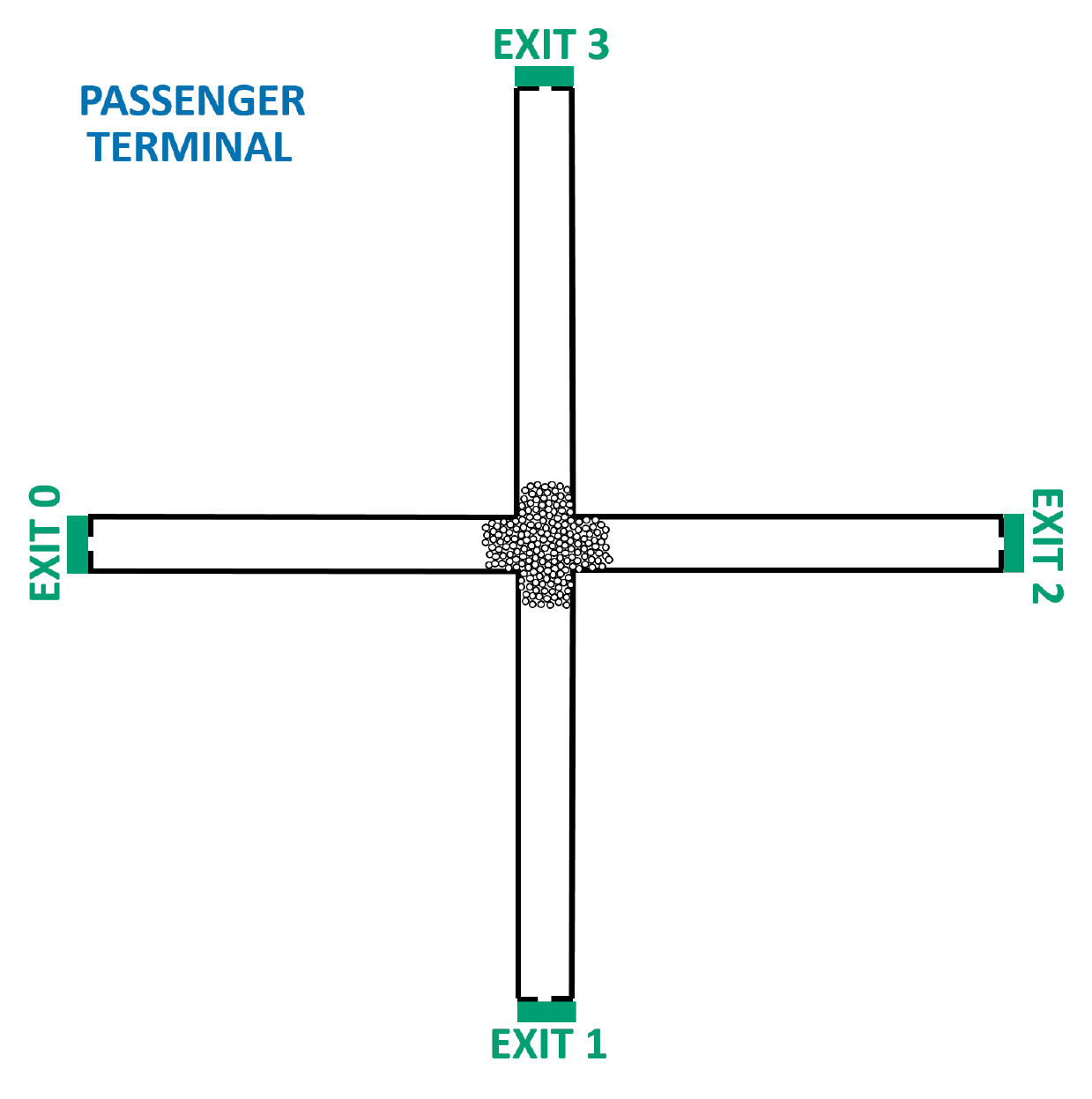}}
\subfigure[]{\includegraphics[width=0.49\textwidth]{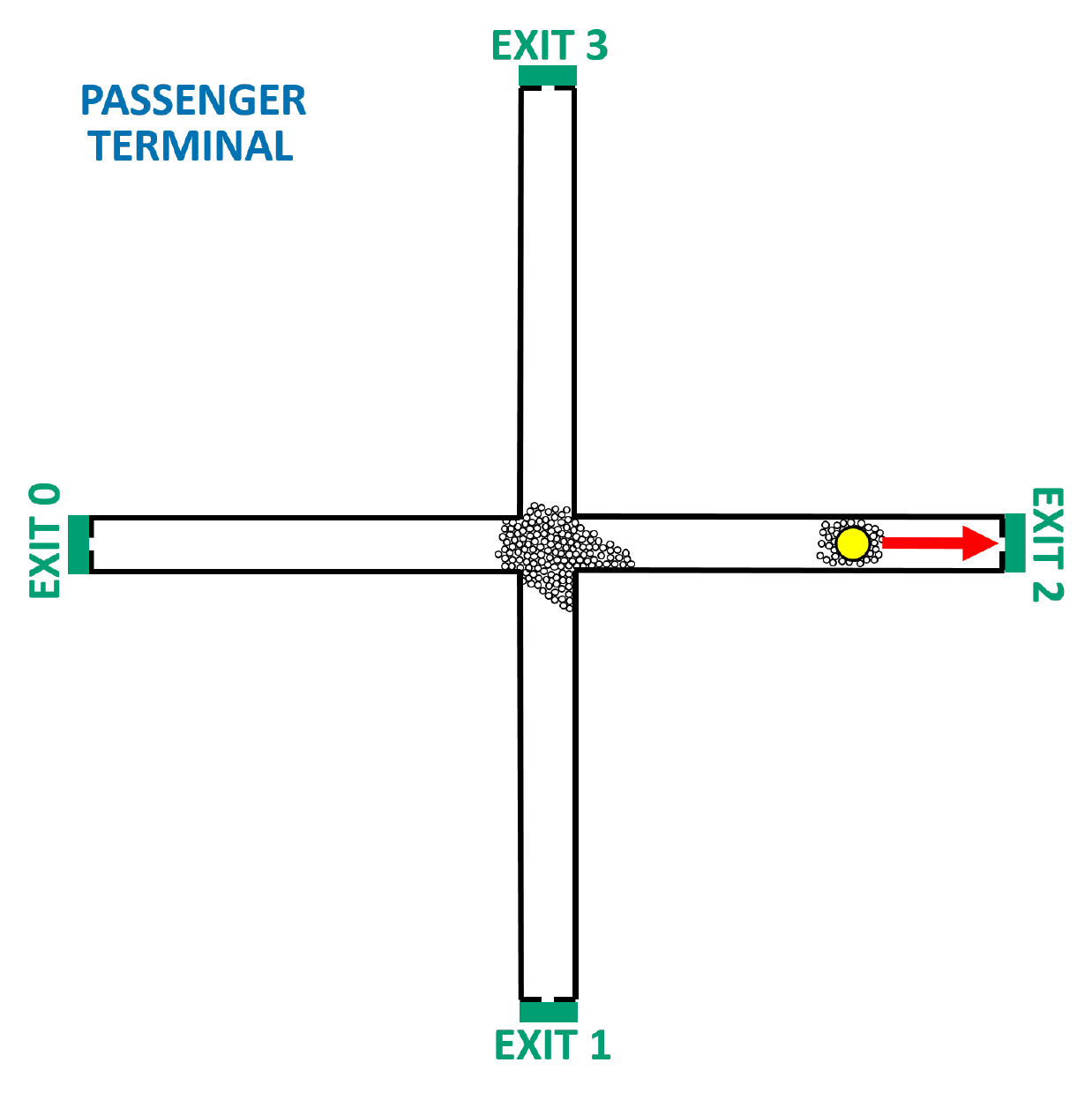}}

\subfigure[]{\includegraphics[width=0.49\textwidth]{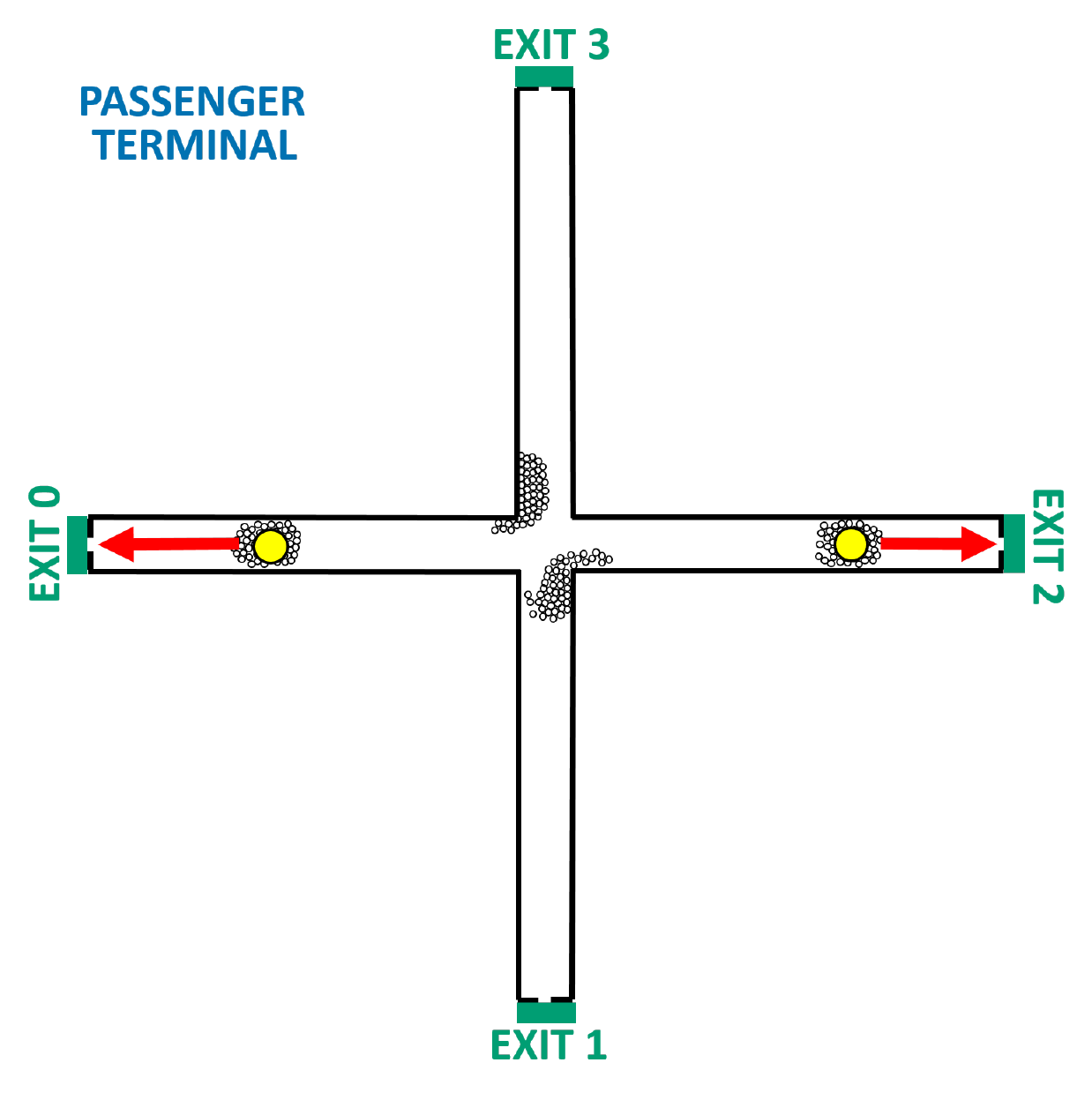}}
\subfigure[]{\includegraphics[width=0.49\textwidth]{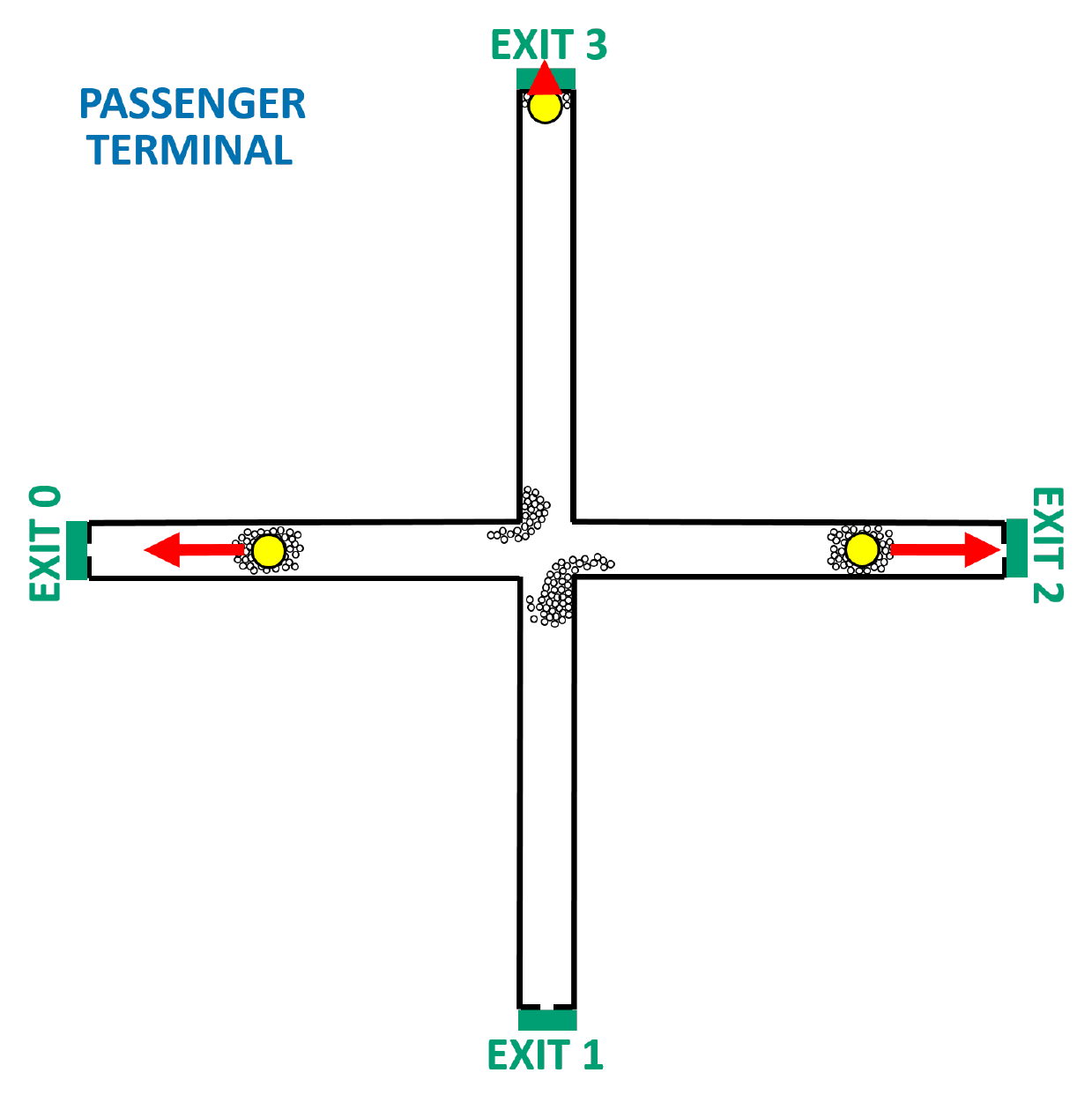}}
\caption{Snapshots of the worst-case scenario (Scenario 2) with $\textrm{CVaR}_{\alpha}$-optimal evacuation plans: (a) $1$ guide, (b) $2$ guides, (c) $3$ guides, and (d) $4$ guides. The snapshots are taken $50$ s from the beginning of the evacuation. The guides' interaction ranges are not drawn in this scheme.}
\label{fig:guides_clogging}
\end{figure}

Let us analyze the ${\textrm{CVaR}}_{\alpha}$-optimal solutions in the worst-case scenario (Scenario $2$). With $0$ guides, the agents move to the exits on the opposite ends of the hallways. At the intersection, they create a congestion (see Fig.~\ref{fig:guides_clogging}(a)). With $1$ guide, the guide starts from the lower hallway and takes a large part of the lower group to Exit $2$. On the way, it encounters the right group and reroutes it to Exit $2$. Still, a small part of the lower group, the left, and the upper group create a congestion at the intersection (see Fig.~\ref{fig:guides_clogging}(b)).

With $2$ guides, the congestion at the intersection is completely cleared. As with $1$ guide, one of the guides starts from the lower hallway and takes a large part of the lower group to Exit $2$. The guide also reroutes the right group to Exit $2$. The second guide starts from the upper hallway and takes the upper group to Exit $0$. The guide also reroutes the right group to Exit $0$. The upper and lower groups do not collide since they move along the walls at the intersection (see Fig.~\ref{fig:guides_clogging}(c)).

With $3$ guides, we notice the same features as with $2$ guides. Also, there is the third guide positioned in the upper hall. It influences part of the upper group members and takes them to Exit $3$. This results in fewer agents going to Exit $0$, which improves the evacuation slightly (see Fig.\ref{fig:guides_clogging}(d)). As we saw in Fig.~\ref{fig:4guides}, with $4$ guides, the agents are all taken to their nearest exits.

\begin{figure}[htb!]
  \centering
  \includegraphics[width=1\textwidth]{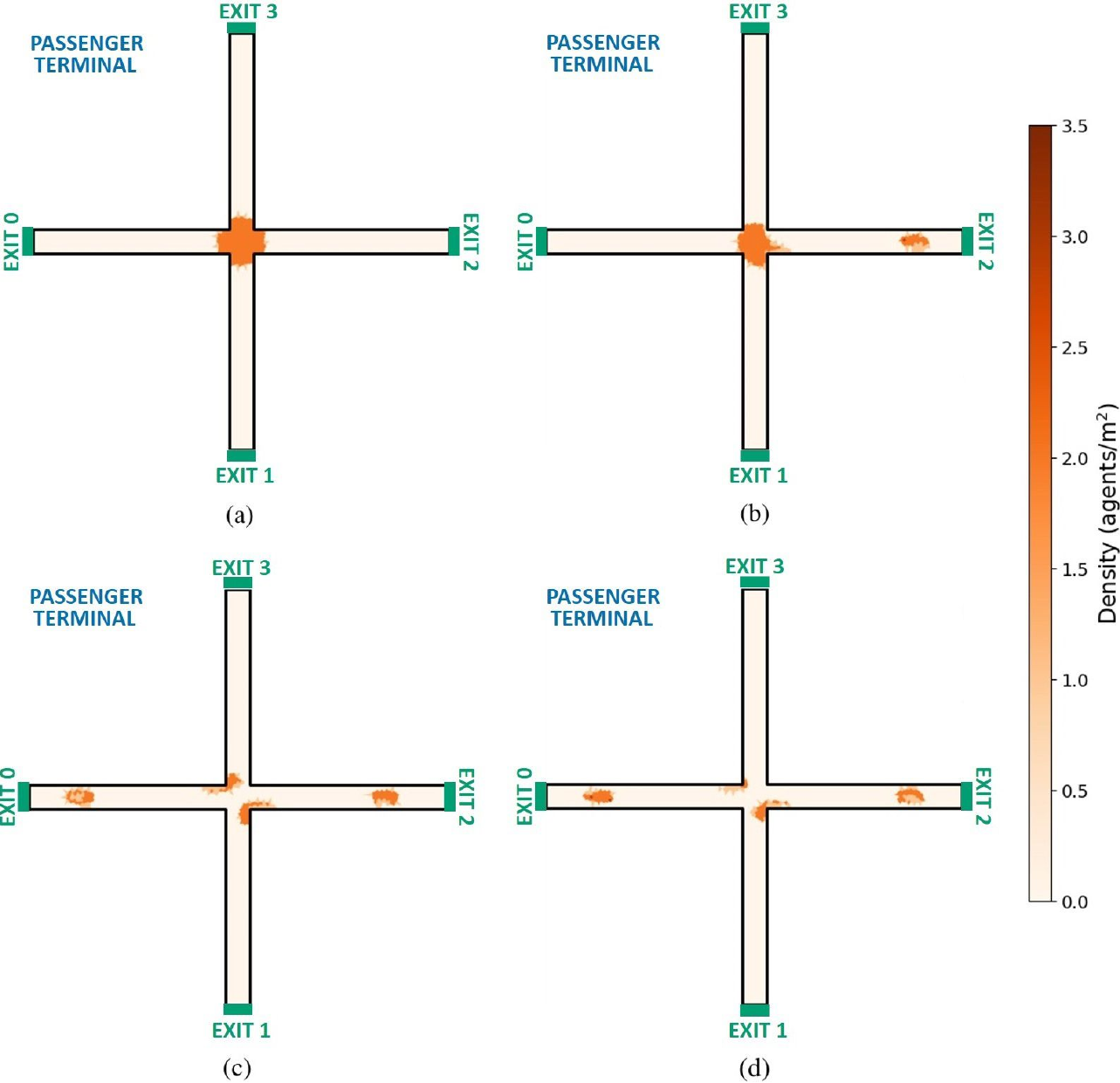}
  \caption{\protect{Density plots of the worst-case scenario (Scenario $2$) with $\textrm{CVaR}_{\alpha}$-optimal evacuation plans: (a) $0$ guides, (b) $1$ guide, (c) $2$ guides, and (d) $3$ guides. The instantaneous densities are calculated $50$ s from the beginning of the evacuation.}}
  \label{fig:guides_density}%
\end{figure}

Another important evacuation performance measure, which is not used as an optimization objective in this paper, is the crowd density. Sometimes the evacuation happens fast enough, but it still is not safe because the crowd density is high. In our example, the density gets largest when the crowd is jammed at the intersection. In \mbox{Fig.~\protect{\ref{fig:guides_density}}} the density of at the intersection is calculated for the same scenario and evacuation plans as in \mbox{Fig.~\protect{\ref{fig:guides_clogging}}} using the detailed method presented in 
\mbox{\protect{\citep{steffen2010methods}}}.

As can be seen, the density does not really exceed $3$ $\text{agents}/\text{m}^2$. In our simulations, the density actually got the highest in Scenario $3$, where the agents run in the same directions as in Scenario $2$ but faster. There, densities of even $5$ $\text{agents}/\text{m}^2$ are observed at the intersection. Such densities in real-life can be dangerous, as people face the risk of falling \mbox{\protect{\citep{smith1995density}}}. High crowd densities could be accounted for in our optimization problem, for example, by adding a penalty if the density exceeds a dangerous level.
\newpage

\newpage

\subsection{Comparison of evacuation times}

\begin{figure}[ht!]
\centering
\subfigure[]{\includegraphics[width=0.45\linewidth]{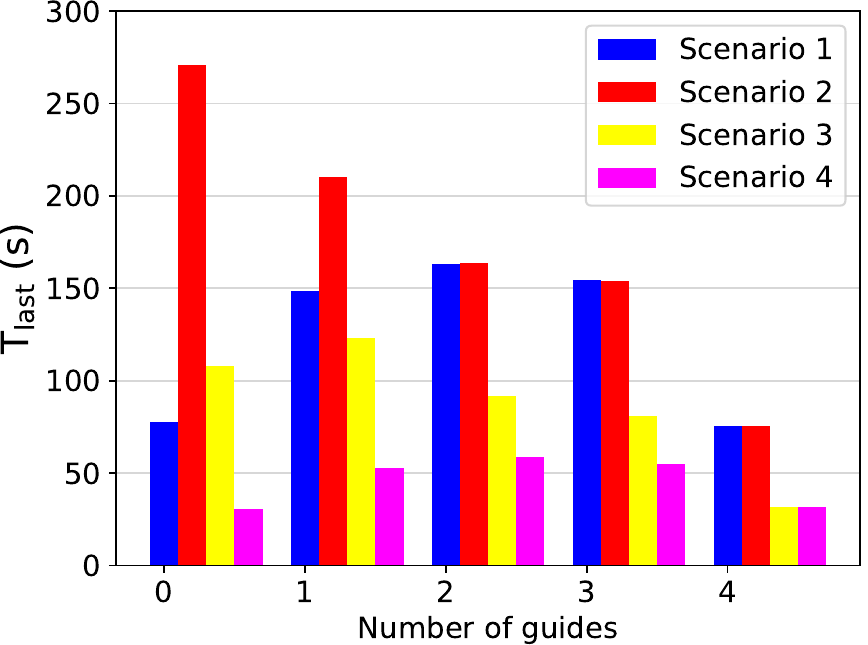}}
\subfigure[]{\includegraphics[width=0.45\linewidth]{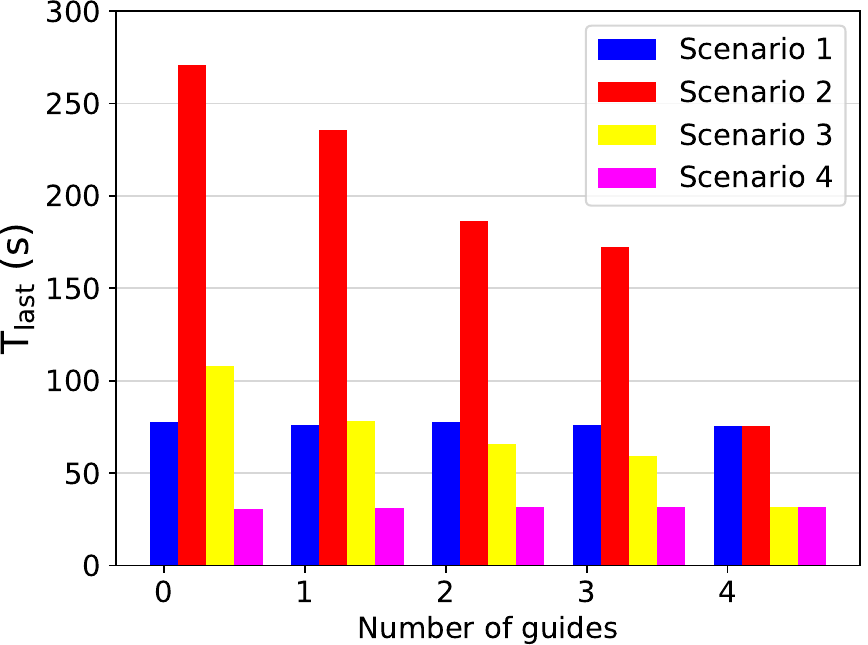}}
\caption{$T_{last}$ (evacuation time) under different scenarios: (a) ${\textrm{CVaR}}_{\alpha}$-optimal solutions, (b) mean-optimal solutions.}
\label{fig:barcharts}
\end{figure}

In Fig.~\ref{fig:barcharts} we see the evacuation times of different scenarios for ${\textrm{CVaR}}_{\alpha}$- and mean-optimal solutions. First, notice that adding guides always decreases evacuation time for Scenario $2$. It is the worst-case scenario, so decreasing its evacuation time improves both objective function values. For mean-optimal solutions, Scenarios $1$ and $4$ are unaffected by an increase in the number of guides. For mean-optimal solutions, the evacuation time of Scenario $3$ is decreased for each added guide.

For ${\textrm{CVaR}}_{\alpha}$-optimal solutions, the results are not as straightforward as for mean-optimal solutions. Using only $1$ guide worsens the evacuation time of Scenario $3$. The reason being that the guide is slower than the passengers, and it reaches the intersection when all agents are jammed there, and influences them all to go to Exit $2$, and creates a further jam there. Also, the evacuation times of Scenarios $1$ and $4$ get worse from adding guides and are the slowest with $2$ guides. With $4$ guides, the evacuation times are again the same as with $0$ guides. The reason for this is quite apparent: the agents evacuate optimally in Scenarios $1$ and $4$ already without any guides, so adding guides elsewhere than close to exits will worsen the evacuation time.

When we monitored the simulations closely, we noticed that on a Pareto front, when moving from the ${\textrm{CVaR}}_{\alpha}$- to the mean-optimal solution, the evacuation time of Scenarios $1$ and $4$ is improved, and the evacuation time of Scenario $2$ is worsened. For Scenario $3$, the optimum would be a solution that is between the ${\textrm{CVaR}}_{\alpha}$-, and mean-optimal solution on the Pareto front.

\subsection{Effect of guides' parameters}

Next, we qualitatively discuss the effect of altering guides' parameters on the Pareto-optimal solutions when all other parameters are held constant. The guides' initial positions can be altered without worsening the solution as long as the interaction range reaches the same passenger agents. All other possible ways the guides' initial positions and exit assignments can be altered are found by rotating $90$, $180$, and $270$ degrees clockwise the diagrams depicting near-optimal solutions (\mbox{Figs.~\protect{\ref{fig:4guides}}, ~\protect{\ref{fig:1guide}}-\protect{\ref{fig:3guides}}}) and by rotating $180$ degrees around the horizontal, vertical, diagonal and cross-diagonal axes. If the guides' initial positions and exit assignments are altered otherwise, the solution gets worse.

Increasing guide speed improves the solution, as faster guides can inform more agents about the optimal evacuation routes. However, it is not their absolute speed that matters, but the relative speed with respect to passenger agents. In our toy example, with faster guides, the $\textrm{CVaR}_{\alpha}$-optimal evacuation plans (\mbox{Figs.~\protect{\ref{fig:1guide}}(a), \protect{\ref{fig:2guides}}(a), and \protect{\ref{fig:3guides}}(a)}) would be optimal for both Scenarios $2$  (slow passenger agents) and $3$ (fast passenger agents). Now they are only optimal for Scenario $2$. In a more sophisticated model, guide speed could also be included as an optimization variable. In a practical implementation, it might be better for the evacuation planner to instruct the guides to finish their route within a time limit rather than making the guides walk at a certain speed.

The guides' interaction range can be thought to be a part of the guiding strategy, e.g., which passenger agents do the guide command to follow him. Increasing the interaction range has mixed effects. Generally, for the near-optimal solutions, it worsens them. On the other hand, if the interaction range is decreased, the evacuation can still be as fast as before changing the parameters. However, the evacuation is then more sensitive to changes in the guides' initial positions.

\subsection{Effect of other model parameters}

Let us briefly discuss how the model parameters affect the Pareto-optimal solutions. We refer to our study \mbox{\protect{\citep{heliovaara2012counterflow}}} for a comprehensive numerical and experimental investigation on the effect of the physical model parameters on congestion in an intersection.

The larger the area the agents are distributed over, or the more sparse the crowd is, the more guides are needed to control it. Nevertheless, with a sparser crowd, physical contact forces are reduced, which results in less congestion and queueing time. A more densely distributed crowd has the opposite effects.
 
In our toy example, the size and initial positions of the crowd are symmetrical. Introducing asymmetry, in most cases, does not affect the optimal evacuation plan notably. Nevertheless, if some groups have more agents or are initially positioned closer to the intersection, they contribute more to the emergence of the congestion. Hence, guides are initially positioned next to these agent groups to steer them away from colliding with other groups.

\section{Discussion and Conclusion}\label{sec:conclusion}

In this paper, we model the movement of a crowd consisting of passengers and guides with a modified social force model. A guide follows routes instructed by the evacuation planner. A passenger goes about its business unless a guide comes within an interaction range and leads it to an exit. The uncertainty of the crowd conditions is described as probabilistic scenarios, which are modeled by altering some of the model input parameters. We formulate the problem of minimizing crowd evacuation time under different scenarios using rescue guides as a bi-objective scenario optimization problem. The two objectives are the mean and ${\textrm{CVaR}}_{\alpha}$ of evacuation time over the scenarios. We present a solution procedure combining numerical simulation and the NSGA-II algorithm. It returns the Pareto-optimal evacuation plans. The procedure is applied to a toy example that is an evacuation of a fictional passenger terminal.

This paper's research question is what happens to the optimal evacuation plan if a large part of a crowd can deviate from its usual behavior. In our example case, for a fixed number of guides less than four, there is a tradeoff between minimizing the mean evacuation time or ${\textrm{CVaR}}_{\alpha}$. With four guides, we obtain a single solution that is optimal for all scenarios.

Often in studies and federal guidelines, evacuation plans are presented by mainly stating the optimal proportion of guides \mbox{\protect{\citep{multigridmodel, leadershipeffect, leadershipeffect2, centripetaleffect, optimizingproportion, dualeffects}}}. We do not think such rules guarantee an efficient evacuation other than in very simple situations. Rather, the optimal evacuation plan is a function of the building geometry \mbox{\protect{\citep{still2000}}}, the initial crowd distribution and behavioral conditions. As could be seen from the optimal evacuation plans in our toy example, with the same number of guides but different route choices, different objectives are optimized. 

In \citep{evacuationassistants} it was found that near-exit positions are good, and in \citep{leadershipeffect} that for building geometries with multiple exits, the guides slow down the evacuation unless all exits are utilized. These findings coincide with our results, given that average performance over scenarios is optimized. Our mean-optimal solutions use a near-exit strategy, where guides are positioned near their destination exits. Maybe a more useful quantity than the initial position is the proximity of a guide to the crowd. In \citep{aubeandshield} it was found that simultaneously positioning guides inside the crowd, on its periphery, and at a distance from it improved evacuation time. However, in the study, the aim was to evacuate a large crowd using the same route. In our study, since the agent groups are relatively small, it is enough to locate one guide per group within the interaction range.

If there is a possibility for a large congestion, and we decide to optimize worst-case performance, a different strategy is needed, as our ${\textrm{CVaR}}_{\alpha}$-optimal solutions show. The guide's focus should be on solving the congestion by moving parts of the crowd elsewhere. Most preferably, the crowd members should be moved away from the congestion area already before the congestion occurs.

In our study, we minimize the evacuation time. It could be interesting to see how the optimal evacuation plans change if there is a time limit within which the crowd should be evacuated. For example, in fire safety literature, a distinction is made between required safe egress time (RSET) and available safe egress time (ASET). RSET defines the time it takes to evacuate the crowd, and ASET the time before the conditions become lethal \mbox{\protect{\citep{ownphysreve, ownphysica}}}. If RSET is less than ASET, the evacuation is efficient, and the evacuation is unacceptable if RSET is more than ASET. The time limit could be implemented into our optimization framework by penalizing for the evacuation time exceeding ASET. So, we would still minimize evacuation time, and at the same time penalize for the evacuation time exceeding ASET.

The framework we have presented here, with some modifications, could also be used for a phased evacuation, i.e., an evacuation where the whole crowd does not exit the building simultaneously. We could implement this by adding the time periods when guides stop their movement as optimization variables. When the guide does not move, neither would the agents that follow it.

In future research, it would be challenging to try our model on a real-world case with real data. If one wants to add more complex crowd behaviors in our framework, it can be done by modeling them as scenarios. If assigning probabilities to the scenarios is difficult, the worst-case performance, in the sense of robust optimization, can be used \mbox{\protect{\citep{robustsurvey}}}. If the crowd's size increases, the numerical simulations with the social force model might become very slow even on a supercomputer due to nonlinear contact forces between agents. We could then deal with the computational inefficiency by performing numerical simulation with an implicit integration scheme \mbox{\protect{\citep{karamouzas2017implicit}}}. Since many buildings and evacuations share similar features, our solution process's efficiency could be further increased. We could solve the optimal evacuation plans for typical situations in advance and store them. A neural network could then be trained by the plans and used to give fast approximate optimal evacuation plans for previously unsolved problems.

\section*{Acknowledgements}

This study was funded by a grant from the Finnish Science Foundation for Technology and Economics. The calculations in this study were performed using computer resources within the Aalto University School of Science "Science-IT" project. We wish to thank our summer assistant Jaan Tollander de Balsch, who considerably helped us develop the simulation codes.


\bibliographystyle{cdbibstyle} 
\bibliography{main} 

\end{document}